\documentclass[12pt]{article}
\usepackage{cite}
\usepackage{pstricks,pstcol} 
\usepackage{graphicx}
\usepackage{amsmath}    
\usepackage{amssymb}
\usepackage{rotating}
\usepackage{epsfig}
\usepackage{bm}       
\usepackage{url}      

\newcommand{\be}{\begin{equation}}
\newcommand{\ee}{\end{equation}}
\newcommand{\bea}{\begin{eqnarray}}
\newcommand{\eea}{\end{eqnarray}}
\newcommand{\dis}{\displaystyle}

\newcommand{\cO}{{\cal O}}

\newcommand{\KLpienu}{\mbox{$K_{L}\to\pi^{\pm}e^{\mp}\nu$}}

\newcommand{\KLpimunu}{\mbox{$K_{L}\to\pi^{\pm}{\mu}^{\mp}\nu$}}
\newcommand{\Rzz}{\mbox{$\Gamma(K_{L}\rightarrow \pi^0\pi^0)$}}
\newcommand{\Rzzz}{\mbox{$\Gamma(K_{L}\rightarrow \pi^0\pi^0\pi^0)$}}

\newcommand{\Rpm}{\mbox{$\Gamma(K_{L}\rightarrow \pi^+\pi^-)$}}
\newcommand{\Rpienu}{\mbox{$\Gamma(\KLpienu)$}}
\newcommand{\Rpimunu}{\mbox{$\Gamma(\KLpimunu)$}}
\newcommand{\Rppp}{\mbox{$\Gamma(K_{L}\rightarrow \pi^+\pi^-\pi^0)$}}

\newcommand{\beq}{\begin{equation}}
\newcommand{\eeq}{\end{equation}}
\newcommand{\bqa}{\begin{eqnarray}}
\newcommand{\eqa}{\end{eqnarray}}

\newcommand{\K}{\mbox{$K$}}

\newcommand{\KL}{\mbox{$K_{L}$}}
\newcommand{\KS}{\mbox{$K_{S}$}}
\newcommand{\KP}{\mbox{$K^{+}$}}

\def\kpi0{K_{L}\to 3\pi^0}
\def\ke3{K_{L}\to\pi^{\pm}e^{\mp}\nu}
\def\k2pi{K_{L} \to \pi^+\pi^-}

\def\kmuii{{K^{+}\to\mu^{+}\nu(\gamma)}}

\newcommand{\KLpm}{\mbox{$K_{L}\rightarrow\pi^{+}\pi^{-}$}}

\newcommand{\KLzz}{\mbox{$K_{L}\rightarrow\pi^{0}\pi^{0}$}}

\newcommand{\KLzzz}{\mbox{$K_{L}\rightarrow \pi^{0}\pi^{0}\pi^{0}$}}
\newcommand{\KLpmz}{\mbox{$K_{L}\rightarrow \pi^{+}\pi^{-}\pi^{0}$}}

\newcommand{\etal}{{\em et al.}}
\def\gev{\,\mathrm{Ge\kern-0.09em V}}
\def\mev{\,\mathrm{Me\kern-0.09em V}}
\def\qsqmax{q^2_\mathrm{max}}
\def\fds{f_{D_s}}
\def\fd{f_D}
\begin{document}

\begin{flushright}
November 2005 
\end{flushright}
\thispagestyle{empty}
\setcounter{page}{0}

\vskip  0.5 true cm 

\begin{center}
{\Large \textbf{Status of the Cabibbo Angle}} \\ [5 pt]
 \textbf{[Proceedings of CKM 2005 - WG 1]} \\  [20 pt]
\textsc{E.~Blucher,${}^{1}$ E.~De~Lucia,${}^{2}$ G.~Isidori,${}^{2}$ V.~Lubicz~${}^{3}$} [conveners] \\
\textsc{H.~Abele,${}^{4}$ V.~Cirigliano,${}^{5}$ R.~Flores-Mendieta,${}^{6}$ J.~Flynn,${}^{7}$ 
 C.~Gatti,${}^{2}$ A.~Manohar,${}^{8}$  W.~Marciano,${}^{9}$   V.~Pavlunin,${}^{10}$   
 D.~Pocanic,${}^{11}$  F.~Schwab,${}^{12}$ A.~Sirlin,${}^{13}$
 C.~Tarantino,${}^{3}$  M.~Velasco~${}^{14}$ }  \\ [20 pt]
${}^{1}~$\textsl{Dept. of Physics, University of Chicago, IL 60637-1434, USA} \\ [5 pt]
${}^{2}~$\textsl{INFN, Laboratori Nazionali di Frascati, I-00044 Frascati,
      Italy} \\ [5 pt]
${}^{3}~$\textsl{ Dipartimento di Fisica, Universit\'a di Roma Tre, and INFN,
      \\  Sezione di Roma Tre, Via della Vasca Navale 84, I-00146 Rome, Italy}  \\ [5 pt]
${}^{4}~$\textsl{Physikalisches Institut der Universit\"at,\\ Philosophenweg 12,
69120 Heidelberg, Germany } \\ [5 pt]
${}^{5}~$\textsl{California Institute of Technology, Pasadena, CA 91125, USA} \\  [5 pt]
${}^{6}~$\textsl{Instituto de Fisica, UASLP, San Luis Potosi 78000 Mexico } \\ [5 pt]
${}^{7}~$\textsl{School of Physics and Astronomy,
Univ. of Southampton, Southampton S017 1BJ, UK} \\ [5 pt]
${}^{8}~$\textsl{Dept. of Phys.,  University of California, San Diego,
La Jolla, CA 92093 USA} \\ [5 pt]
${}^{9}~$\textsl{     Brookhaven National Lab., PO Box 5000, Upton, NY 11973-5000, USA
} \\ [5 pt]
${}^{10}~$\textsl{Dept. of Physic,  Purdue University, \\525 Northwestern Avenue,
West Lafayette, IN 47907, USA
    } \\ [5 pt]
${}^{11}~$\textsl{Dept. of Physics, University of Virginia,\\ PO Box 400714
Charlottesville, VA 22904-4714 USA
    } \\ [5 pt]
${}^{12}~$\textsl{ Physik-Department, Technische Universitaet Muenchen, D-85748 Garching,\\
Max-Planck-Institut fuer Physik, Foehringer Ring 6, D-80805 Muenchen, Germany } \\ [5 pt]
${}^{13}~$\textsl{Dept. of Physics, New York University,
New York, NY 10003, USA
      } \\ [5 pt]
${}^{14}~$\textsl{ Dept. of Physics and Astronomy, Northwestern University, \\
Evanston, Illinois 60208-3112    } \\ [5 pt]

\vskip   1  true cm 

\textbf{Abstract}\\
\end{center}
\noindent 
We review the recent experimental and theoretical progress in 
the determination of $|V_{ud}|$ and $|V_{us}|$, 
and the status of the most stringent test of CKM unitarity.
Future prospects on $|V_{cd}|$ and $|V_{cs}|$ are also briefly discussed.

\newpage
\tableofcontents      
\newpage
 
\section{Introduction} \label{intro}  

The most precise constraints on the size of the elements of the 
CKM matrix~\cite{Cabibbo,KM} are extracted from the low-energy  
$s\rightarrow u$ and $d\rightarrow u$ 
semileptonic transitions. Combining the precise determinations of
$|V_{ud}|$ and $|V_{us}|$ extracted from these processes 
we can perform the most stringent test of CKM unitarity, namely we can
probe the validity of the relation 
\begin{equation}
|V_{ud}|^2+|V_{us}|^2+|V_{ub}|^2 =1~.
\label{eq:unitarity}
\end{equation}
at the  0.1\% level. In particular, the best determination of $|V_{us}|$ 
is obtained from semileptonic $K$ decays ($K_{\ell3}$ and $K_{\ell2}$),
while the most stringent constraints on $|V_{ud}|$ are
obtained from superallowed Fermi transitions (and, to a minor extent, 
from neutron and pion beta decays). 
As we will discuss in the following, in the last 
few years there has been a substantial progress
\-- both from the theoretical and the experimental side \--
in  the determination of these two matrix elements.

In all cases the key observation which allow a precise 
extraction of the CKM factors is the conservation
of the vector current at zero momentum transfer
in the $SU(N)$ limit and the non-renormalization theorem.  
The latter implies that the relevant hadronic
form factors are completely determined up to tiny second order
isospin-breaking corrections in the $d\rightarrow u$ case~\cite{behr} 
or SU(3)-breaking corrections in the $s\rightarrow u$ case~\cite{ag}.
As a result of this fortunate situation, the accuracy on
$|V_{us}|$ is approaching the 1\% level and the one on 
$|V_{ud}|$ is already below the 0.05\% level. 

The present accuracy on $|V_{ud}|$ and $|V_{us}|$ is such 
that the contribution of $|V_{ub}|^2\approx 2 \times 10^{-5}$ 
in the relation (\ref{eq:unitarity}) can safely be neglected, 
and the uncertainty of the first two terms is comparable. 
In other words, to a high degree of accuracy we can set 
\begin{equation}
V_{us} = \sin\theta_c~,\qquad   V_{ud} = \cos\theta_c~,
\end{equation}
as in the original Cabibbo theory~\cite{Cabibbo},
and $|V_{us}|$ and  $|V_{ud}|$ provide 
two independent determinations of the Cabibbo angle 
both around the 1\% level.

In the following sections we will review the 
determinations of $|V_{us}|$ and $|V_{ud}|$ from 
the main observables mentioned above.
We will also briefly analyze alternative strategies 
to extract  $|V_{us}|$ from $\tau$ and hyperon decays,
as well as the future prospects on $|V_{cs}|$ and $|V_{cd}|$.
The main recent results on $|V_{us}|$ and $|V_{ud}|$
are then summarized and combined in the last section,
where we will discuss the accuracy to which  Eq.~(\ref{eq:unitarity})
is satisfied and we will provide a final global 
estimate of the Cabibbo angle.

\section{The extraction of $V_{ud}$}
\label{sect:Vud}

The value of $V_{ud}$ has been extracted from 1) super-allowed,
$0^{+}\rightarrow0^{+}$, nuclear beta decays, 2) neutron beta decays,
$n \rightarrow pe\nu$, and 3) pion beta decay
$\pi^{+}\rightarrow\pi^{0}e^{+}\nu$.
The latter two, subsequently 
discussed in this report, have smaller overall theoretical
uncertainties and may in the long term be better ways to obtain
$V_{ud}$; but currently, only super-allowed beta decays determine
$V_{ud}$ to better than 0.05\%.

\subsection{Super-allowed Fermi transitions}   
\label{sect:Vud_SFT}

The so-called super-allowed, $0^{+}\rightarrow0^{+}$, Fermi
transitions between nuclei are very special~\cite{sirlin}. Because
they proceed (at the tree level) through pure weak vector current
interactions, which are conserved in the $m_{d}=m_{u}$ limit; they are
not renormalized by strong interactions at $q^{2}=0$. Hence, they are
ideally suited for cleanly extracting $V_{ud}$ with high
precision. Corrections due to $q^{2}\neq 0$ and $m_{d}\neq m _{u}$ are
negligibly small; so, one needs only to control uncertainties in the
electroweak radiative corrections, isospin violating electromagnetic
effects and nuclear structure dependence. How well that can be done is
the subject of this section.

Last year, the prevailing value of $V_{ud}$ obtained by averaging the
nine best measured super-allowed $\beta$-decays was~\cite{czarnecki,towner}
\begin{equation}
V_{ud}=0.9740(1)(3)(4)\rightarrow0.9740(5) \hspace{1cm}     
{\rm (2004\ value)}
\end{equation}
where the errors are experimental, nuclear theory and quantum loop
corrections. The very small experimental error illustrates the power
of this averaging procedure. The largest uncertainty, associated with
weak axial-vector induced loop effects~\cite{marciano12}
primarily through $\gamma W$ box diagrams, represents model dependent
hadronic effects which until recently~\cite{marciano13} were thought
to be essentially irreducible or at least very difficult to reduce.

Two developments have led to a recent improvement in $V_{ud}$ by
nearly a factor of 2. First, new global studies of super-allowed
$\beta$-decays by Hardy and Towner~\cite{hardy}, and 
by Savard {\em et al.} \cite{savard}, have provided a more
consistent treatment of $Q$ values and lifetimes used in $ft$
determinations, which in turn give $V_{ud}$ via the master formula
\begin{equation}
|V_{ud}|^{2}=\frac{2984.48(5)~{\rm s}}{ft(1+RC)}              
\end{equation}
In that expression, RC designates the total effect of all radiative
corrections from quantum loops as well as nuclear structure and
isospin violating effects. RC is nucleus dependent, ranging from about
+3.1\% to +3.6\% for the nine best measured super-allowed decays. That
difference is of critical importance in bringing the values of
$V_{ud}$ obtained from separate decays into agreement with one
another. The magnitude of the corrections is essential for
establishing unitarity, as we shall see.

A second major advance in the determination of $V_{ud}$ stems from a
new study of the quantum loop corrections coming from the 
problematic $\gamma {\rm W}$ box diagram due to weak axial-vector
contributions. Previously, those effects, along with other smaller
axial-vector current contributions, were found to shift the RC by
about
\begin{equation}
\frac{\alpha}{2\pi}\biggl[ \ln\frac{m_{Z}}{m_{A}}+A_{g}+2C_{\rm Born}\biggr]
\end{equation}                                         
where $A_{g}={- 0.34}$ is a one-loop QCD correction to the short-distance
logarithmic loop contribution and
$C_{Born}\simeq0.8g_{A}(\mu_{p}+\mu_{n})\simeq0.9$ represents
long-distance loop effects. The problematic intermediate loop momentum
region was roughly estimated by employing $m_{A}\simeq1.2{\rm GeV}$ in
the log, while the crudely obtained error of $\pm8\times10^{-4}$ in
that quantity (which leads to $\pm4\times10^{-4}$ in $V_{ud}$) was
found~\cite{czarnecki, marciano12} by allowing the $m_{A}$ cut-off
scale to vary up or down by a factor of 2.

A new analysis~\cite{marciano13} of the $\gamma{\rm W}$ box diagram
now divides the loop momentum into 3 integration regions:
\begin{eqnarray}
&(1.5~{\rm GeV})^{2}\le Q^{2}_{I}<\infty& \nonumber\\
&(0.82~{\rm GeV})^{2}\le Q^{2}_{II}<(1.5~{\rm GeV})^{2}& \nonumber \\
&0\le Q^{2}_{III}<(0.82~{\rm GeV})^{2}& \nonumber
\end{eqnarray}

The evaluation of region I has been supplemented by 3-loop QCD
corrections to the leading term in the short-distance operator product
expansion, rendering it effectively error free and, more important,
allowing a smooth extrapolation to lower $Q^{2}$. Region II has
been evaluated using interpolating vector and axial-vector resonances,
a procedure motivated by large $N_{c}$ QCD and vector meson
dominance. That prescription has been well tested in other
calculations; nevertheless, a conservative $\pm100\%$ uncertainty has
been assigned to that part of the calculation. Finally, region III was
evaluated using well-measured nucleon dipole form factors and assigned
a $\pm10\%$ uncertainty. Those improvements have reduced the
theoretical quantum loop uncertainty in $V_{ud}$ from a crude
$\pm4\times10^{-4}$ to a more defensible $\pm1.9\times10^{-4}$, about a 
factor of 2 improvement. Further error
reduction may be possible if future lattice calculations can confirm
the interpolating resonance approach, since the uncertainty from
intermediate momenta is still dominant.

The overall shift in $V_{ud}$ due to the new evaluation of radiative
corrections is relatively small. However, the
error reduction is more significant. Updating the most recent 
$ft$ values~\cite{savard} with the new RC results leads to the
$V_{ud}$ values given in Table~1. Combining all errors in quadrature
now gives the weighted average~\cite{marciano13}
\begin{equation}
V_{ud}=  0.97377(27)  \hspace{1cm} (\rm 2005\ value)   
\label{eq:Vud2005}
\end{equation}
The central value has not shifted very much [see Eq.~(7)], but the error
has been reduced by nearly a factor of 2.

\begin{table}[t]
\begin{center}
\begin{tabular}{|c|c|c|c|c|}  \hline
Nucleus & $ft$\ (sec) &  $V_{ud}$\\ \hline  \hline  
$^{10}$C & 3039.5(47) &  0.97381(77)(15)(19)\\
$^{14}$0 & 3043.3(19) &  0.97368(39)(15)(19)\\
$^{26}$Al & 3036.8(11) & 0.97406(23)(15)(19)\\
$^{34}$Cl & 3050.0(12) &  0.97412(26)(15)(19)\\
$^{38}$K & 3051.1(10) &  0.97404(26)(15)(19)\\
$^{42}$Sc & 3046.8(12) &  0.97330(32)(15)(19)\\
$^{46}$V & 3050.7(12) &  0.97280(34)(15)(19)\\
$^{50}$Mn & 3045.8(16) & 0.97367(41)(15)(19)\\
$^{54}$Co & 3048.4(11) &  0.97373(40)(15)(19)\\ \hline\hline
\multicolumn{2}{|c|}{ weighted ave.} & 0.97377(11)(15)(19)\\
\hline
\end{tabular}
\end{center}
\caption{  Values of $V_{ud}$ implied by various precisely measured
superallowed nuclear beta decays~\cite{marciano13}. The $ft$ values are taken from a
recent update by Savard {\em et al.}~\cite{savard}. Uncertainties in
$V_{ud}$ correspond to 1) nuclear structure and $Z^{2}\alpha^{3}$
uncertainties added in quadrature with the $ft$ error, 2) a common
error assigned to nuclear coulomb distortion effects, and 3) a
recently reduced (common) uncertainty in the radiative corrections
from quantum loop effects. Only the first error is used to obtain the
weighted average. }
\end{table}

So far the situation for $V_{ud}$ looks very good; however, we caution 
that the recent re-measurement of the $Q$ value for $^{46}V$~\cite{savard}
has implied a substantial ($\sim 2 \sigma$) shift in the corresponding 
extraction of  $V_{ud}$ [from $0.97363(50)$ to $0.97280(43)$]. 
As a result, the overall consistency of the  $V_{ud}$ values extracted 
from the various nuclei is not as good as it was about one year ago.
This fact could be interpreted as an indication of possible 
problems with the $Z$-dependent radiative corrections or the $Q$ values 
of other superallowed decays. The shift in the total average of $|V_{ud}|$
implied by the new $^{46}V$ data is not significant, but 
changes in the other superallowed $Q$ values could have more 
substantial effect. 

The superallowed beta decays have now reached the very impressive $\pm
0.03\%$ level of precision in their determination of $V_{ud}$. Further
studies of those reactions are clearly warranted, both to reduce the
error and to clarify the new $^{46}V$ anomaly. In addition, future
high statistics neutron-studies~\cite{czarnecki,HADM} of $\tau _{n}$ and $g_{A}$
may be able to reach a level of precision for $V_{ud}$ comparable to
Eq.~(\ref{eq:Vud2005}), 
but without the nuclear physics complications. Those
measurements, which are difficult but well worth the effort,
will be discussed in the next subsection.

\subsection{Neutron $\beta$-decay }  
The beta decay of the neutron allows a determination of  
$|V_{ud}|$  free from
the nuclear structure effects of superallowed beta decays. 
Fixing the Fermi constant from the muon decay, 
within the Standard Model (SM) we can describe the neutron $\beta$-decay 
in terms of two parameters. One of them is $|V_{ud}|$,
the other is the ratio of axial and vector  coupling constants
relevant to the $n\to p e \nu$ transition: $\lambda = g_A/g_V$.
The determination of $|V_{ud}|$ is then based on two experimental
inputs: the neutron lifetime and $\lambda$.

The neutron lifetime can be written as  
\begin{equation}
\tau_n = \frac{K}{|V_{ud}|^2 G_F^2 (1+3\lambda^2)(1+\Delta_R) f^R } \ \ ,
\label{eq:taun}
\end{equation}
where $f^R = 1.71335(15)$~\cite{Wil82,czarnecki}
is the phase space factor (corrected for the 
model independent radiative corrections), 
$\Delta_R = 0.0239(4)$ denotes the 
model dependent radiative corrections 
to the neutron decay rate \cite{marciano13,Towner1}
and  $K$ is an appropriate normalization constant 
(see e.g.~Ref.~\cite{CKMBook}).

The most precise experimental information on $\lambda$ 
is derived from the $\beta$-asymmetry coefficient $A_0$, 
which describes the correlation between the neutron spin 
and the electron momentum. To a minor extent, 
also the correlation between neutrino and electron momenta 
and the correlation between neutron spin and proton momentum
are sensitive to $\lambda$. The neutron  $\beta$ decay is therefore 
an overconstrained system. In principle, $\lambda$ could  
also be determined from lattice QCD; however, at present 
the results of the most precise calculations 
are affected by ${\cal O}(30\%)$ errors. 

Since the  overall uncertainty in $|V_{ud}|$ is dominated by the 
experimental errors on $\tau_n$ and $A_0$, in the following 
we restrict our discussion on these two main observables.

\subsubsection{First input: $\lambda$ from the $\beta$-asymmetry coefficient $A_0$}

The coefficient $A_0$ is linked to the probability that an electron is
emitted with angle $\vartheta$ with respect to the neutron spin
polarization $P$:
\begin{equation} W(\vartheta) = 1 +\frac{v}{c}PA_0\cos(\vartheta)
\end{equation}
where $v/c$ is the electron velocity expressed in fractions of the
speed of light. Neglecting order 1\% corrections, $A_0$ is a
simple function of $\lambda$:
\begin{equation}  A_0=-2\frac{\lambda(\lambda+1)}{1+3\lambda^2},
\end{equation} where we have assumed that $\lambda$ is real.

The most precise measurement of $A_0$, 
recently obtained by means of the instrument PERKEO,
results in ${\lambda=-1.2739(19)}$~\cite{Abele02}. 
A recent repetition of this experiment confirms this value. 
In this experiment, the total correction to the raw data is 0.4\% 
and the error contribution to $|V_{ud}|$ is $\pm 0.0007$~\cite{Mund}.
Earlier experiments~\cite{Bopp,Yerozolimsky,Schreckenbach} gave
significantly lower values for $A_0$. Averaging over recent experiments 
using polarizations of more than 90\%, the Particle Data 
Group~\cite{PDG2004} obtains the value $\lambda=-1.2720(18)$.

About half a dozen new instruments have been 
planned or are under construction for measurements
of $A_0$ and the other  $\beta$-asymmetry coefficients at the sub-10$^{-3}$ level.
Once this program will be completed, the error on 
$|V_{ud}|$ due to the determination of 
$\lambda$ will be subleading with respect to the 
theoretical uncertainties of radiative corrections.
Better neutron sources, in particular for high fluxes of cold and high
densities of ultra-cold neutrons will boost the fundamental
studies in this field. Major improvements both in neutron flux and
degree of neutron polarization have already been made: first, a
ballistic supermirror guide at the Institute-Laue Langevin in
Grenoble gives an increase of about a factor of 4 in the cold
neutron flux~\cite{Haese}. Second, a new arrangement of two
supermirror polarizers allows to achieve an unprecedented degree
of neutron polarization $P$ of between 99.5\% and 100\% 
over the full cross section of the beam~\cite{Soldner}. 
Future trends have been presented at the workshop
``Quark-mixing, CKM Unitarity"~\cite{HADM} and the ``International
Conference on Precision Measurements with Slow Neutrons"
\cite{PMSN}.

\subsubsection{Second input: the neutron lifetime $\tau_n$}

The world average value for the neutron mean lifetime includes
about a dozen individual measurements, using different techniques
summarized as ``in-beam methods" and ``bottle methods". The ``in-beam
methods" count the neutron decay product near a slow neutron beam
whereas the ``bottle methods" store ultra-cold neutrons and count
the neutrons that survived a certain time interval. One of the key
points of neutron storage experiments is the control of neutron
losses: in all bottle experiments performed so far, corrections of 
several percent are necessary because of the losses of ultra-cold 
neutrons at the material-trap walls. The best beam experiments have 
overall larger errors, but they need to apply smaller corrections.

In 2004, all measurements agreed nicely with a $\chi^2$/d.o.f.=0.95. 
The world average value was dominated by a single experiment~\cite{Arzumanov}
and the value adopted by the Particle Data Group was $\tau_n$ = 885.7(8)s~\cite{PDG2004}.
The error contribution to $|V_{ud}|$ from this value of the
neutron lifetime is $\pm0.0004$, which is subleading with respect 
to the error induced by~$\lambda$.

The situation has changed a few months ago,
after the results of a new bottle experiment
where the storage time for ultra-cold
neutrons is very close to the neutron lifetime. 
The result reported by this experiment is 
$\tau_n =878.5(8)$s~\cite{Serebrov04}, which 
differs by more than 6$\sigma$ from
the Particle Data Group average. Such a change in
the neutron lifetime would  have a very significant effect on 
CKM-unitarity. On the other hand, it is worth noting that this value 
for $\tau_n$ brings the Big Bang nucleo-synthesis scenario in better 
agreement with the independent determination of the baryon content 
of the Universe deduced from the WMAP analysis
of the cosmic microwave background power spectrum~\cite{Mathews}.

A new generation of $\tau_n$-experiments, performed with magnetic 
storage devices where wall contacts of neutrons and thus neutron losses
are avoided, is under construction. In this new approach, neutrons are trapped
magnetically with permanent magnets~\cite{Ezhov} or with the
superconducting magnets~\cite{Hartmann,Huffman} (see also
the proceedings of~\cite{HADM,PMSN}).

\subsubsection{Results}
Inverting Eq.~(\ref{eq:taun}) leads to the following  
master formula~\cite{czarnecki,marciano13}:
\begin{equation}
|V_{ud}|^{2}=\frac{(4908.7 \pm 1.9){\rm s}}{\tau_{n}(1 +3\lambda^2)}~.
\label{4908}
\end{equation}
Employing $\tau_{n} = 885.7(7)$s and $\lambda$ = 1.2720(18) this
implies
\begin{equation}
|V_{ud}| = 0.9730\pm0.0004\pm0.0012\pm \pm0.0002~,
\label{0971}
\end{equation}
where the errors stem from the experimental uncertainty in the
neutron lifetime, the $\beta$-asymmetry $A_0$ and the theoretical
radiative corrections, respectively. As can be noted, so far the error
is dominated by experimental uncertainties. A lower lifetime
$\tau_n$ = 878.5(8)s would change $|V_{ud}|$ to 0.9769(13).

\subsection{$V_{ud}$ from $\pi_\beta$ decays: the PIBETA experiment} 

Pion beta decay, $\pi^+\to\pi^0e^+\nu$ (also labeled $\pi_\beta$ and
$\pi_{\rm e3}$), provides a theoretically exceptionally clean means to
study weak $u$-$d$ quark mixing, i.e., the CKM matrix element
$V_{ud}$.  Recent calculations of radiative corrections in Ref.~\cite{Jau01} 
and Ref.~\cite{Cir03} demonstrate that the
theoretical uncertainty accompanying $V_{ud}$ extraction from pion
beta decay is below 0.05\%. The new theoretical analysis of
Ref.~\cite{marciano13}
indicates a further improvement in theoretical accuracy.
In the past, the low branching ratio of pion beta decay, $B_{\pi\beta}
\sim 10^{-8}$, has stood in the way of fully exploiting this
opportunity experimentally.  Until recently, the most accurate
published measurement of $B_{\pi\beta}$ was the one made by McFarlane
{\em et al.}~\cite{McF85}, with relative uncertainty $\Delta B/B$ barely
under 4\%, not even low enough to test the validity of radiative corrections
 which amount to approximately 3.2\%.

The PIBETA collaboration\footnote{Univ.\ of Virginia, PSI, Arizona
State Univ.\ , JINR Dubna, Swierk, Tbilisi, IRB Zagreb} has initiated
a program of measurements to improve the experimental precision of the
branching ratios of pion beta and other rare pion as well as muon
decays at the Paul Scherrer Institute (PSI) in Switzerland.  Detailed
information on the experimental method is available in Ref.\
\cite{Frl04a} and at {\sl http://pibeta.phys.virginia.edu/}.

\begin{figure}[t]
\hbox to \textwidth{
\resizebox{0.60\textwidth}{!}{\includegraphics[width=5cm]{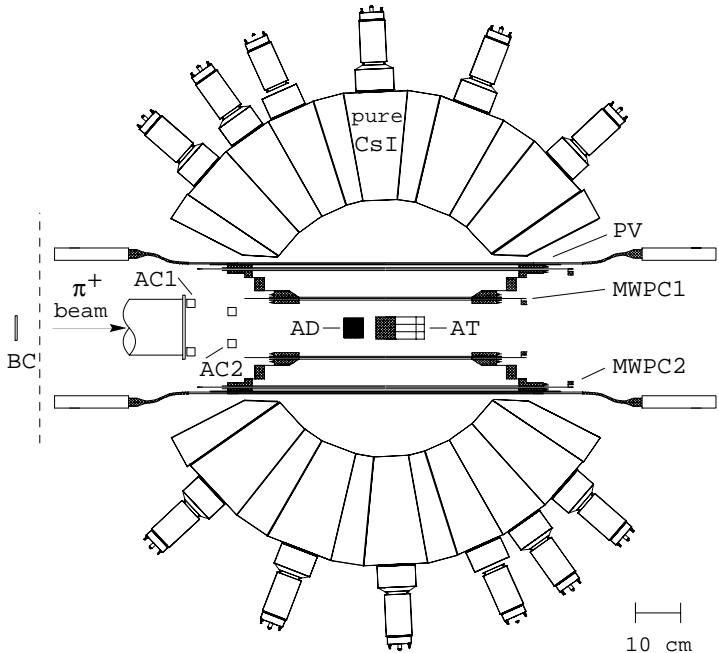}}
\hfil \vbox{\hsize=0.34\textwidth
\caption{Schematic cross section of the PIBETA apparatus showing the
main components: beam coun\-ters (BC, AC1, AC2), ac\-tive degrader (AD) and
target (AT), MWPC's and support, plastic veto (PV) detectors and
PMT's, pure CsI calorimeter and PMT's.
\label{fig:xsect}
       } }}
\end{figure}

The experiment measures decays at rest.  A 114\,MeV/c pion beam is
tagged in an upstream beam counter (BC), slowed down in an active
degrader (AD), and stopped in a segmented 9-element active target
(AT).  Charged particles are tracked in a pair of thin concentric
multiwire proportional chambers (MWPC) and a 20-bar thin plastic
scintillator veto hodoscope (PV).  A large-acceptance 240-element
spherical electromagnetic shower calorimeter made of pure CsI (12
radiation lengths thick) is used to detect the energy of both charged
and neutral particles.  Layout of the experiment is depicted
schematically in Fig.\ \ref{fig:xsect}.  The $\pi^+ \to\rm e^+\nu$
decay events ($\pi_{\rm e2}$) were used for normalization.

The first phase of the experiment ran from 1999 to 2001.  All observed
decays were analyzed in parallel in order to minimize systematic
uncertainties through multiple redundant crosschecks.  A follow-up
run, optimized for the $\pi^+ \to\rm e^+\nu\gamma$ radiative decay,
took place in the summer of 2004.  The main results of the 1999-2001
run were published in Refs.\ \cite{Poc04, Frl04b}.  Because of the
normalization to $\pi_{\rm e2}$ decays, one can evaluate the pion beta
decay branching ratio in two ways.  Assuming full validity of the
Standard Model, we can multiply the measured ratio
$B_{\pi\beta}/B_{\rm\pi e2}$ by the theoretical value of $B^{\rm
SM}_{\rm\pi e2} = 1.2352\,(5) \times 10^{-4}$ from Ref.\ \cite{Mar93}
and obtain (method 1):
\begin{equation}
   B^{\text{exp}}_{\pi\beta} = \rm
       [1.040 \pm 0.004\,(stat) \pm 0.004\,(syst)]
                 \times 10^{-8} \,.   \label{eq:br_theo_norm}
\end{equation}
Alternatively, taking the experimental 
world average for the $\pi_{\rm e2}$ branching ratio, $B^{\rm
exp}_{\rm\pi e2} = 1.230\,(4) \times 10^{-4}$ \cite{PDG2004}, one obtains
(method 2):
\begin{equation}
   B^{\text{exp}}_{\pi\beta} = \rm
       [1.036 \pm 0.004\,(stat) \pm 0.004\,(syst) \pm 0.003\,(\pi_{e2})]
                 \times 10^{-8} \,.   \label{eq:br_expt_norm}
\end{equation}
In either case, the result is in excellent agreement with the 90\%
C.L. SM limits: $B_{\pi\beta} = 1.038 - 1.041 \times 10^{-8}$ 
\cite{PDG2004}.  These
results represent the most stringent test of the CVC hypothesis
in a meson performed so far.  Furthermore, they
lie about 6$\sigma$ outside the
SM limits if radiative corrections are excluded, thus confirming the
latter for the first time.  The value of $V_{ud}$ obtained
from these results:  $V_{ud}=0.9748\,(25)$ (method 1), or 0.9728\,(30) (method 2) 
is not yet competitive, but it agrees well with the one of 
superallowed transitions. We stress that this is not the final result of the
PIBETA collaboration: an update with
improved accuracy will be forthcoming.  However, bringing the error
down by an order of magnitude will require a whole new phase of the
experiment.

In the next stage of the project, the PIBETA collaboration will turn its
attention to a new precise measurement of the $\pi^+ \rm\to e^+\nu$
branching ratio.

\section{$V_{us}$ from $K$ decays}
\label{sect:Kl3_th}

\subsection{Theoretical aspects of $K_{\ell3}$ decays}

The decay rates for all four $K_{\ell 3}$ modes ($K={K^\pm, K^0}, 
\ell=\mu,e$)  can be written 
compactly as follows:
\begin{equation}
\Gamma (K_{\ell 3 [\gamma] })  = \frac{G_F^2 \,  S_{\rm ew} 
\,  M_K^5}{128 \pi^3} 
\, C^K  I^{ K \ell} (  \lambda_i )  \times 
| V_{us} \times f_{+}^{K^0 \pi^-} (0)|^2 \times \Big[ 
1 + 2\, \Delta^K_{SU(2)} + 2\,  \Delta^{K \ell}_{\rm EM} 
\Big] \ . 
\label{eq:masterkl3}
\end{equation}
Here $G_F$ is the Fermi constant as extracted from muon decay,
$S_{\rm ew} = 1 + \frac{2 \alpha}{\pi} \left( 1 -\frac{\alpha_s}{4 \pi} \right)\times 
\log \frac{M_Z}{M_\rho}  + O (\frac{\alpha \alpha_s}{\pi^2})$ 
represents the short distance electroweak correction to semileptonic
charged-current processes,  $C^K$ is a Clebsh-Gordan coefficient equal
to 1 (1/$\sqrt{2}$) for neutral (charged) kaon decay, while $I^{ K \ell} (
\lambda_i )$ is a phase-space integral depending on slope and
curvature of the form factors. The latter are 
defined by the QCD matrix elements
\begin{equation}
\langle \pi^j (p_\pi) | \bar{s} \gamma_\mu u | K^i (p_K) \rangle = 
f_{+}^{K^i \pi^j} (t)  \, (p_K + p_\pi)_\mu  + 
f_{-}^{K^i \pi^j} (t)  \, (p_K - p_\pi)_\mu ~.
\end{equation}
In the physical region these can be conveniently parameterised as 
\begin{eqnarray}
f^{K^i \pi^j}_0 (t) & \equiv & f^{K^i \pi^j}_{+} (t) 
+ \displaystyle\frac{t}{M_K^2 - M_\pi^2} \, f^{K^i \pi^j}_{-} (t) \ ,  \\
f^{K^i \pi^j}_{+,0} (t) &=&  f^{K^i \pi^j}_{+} (0)   \, \left( 1 +  \lambda^\prime_{+,0}  
\ \displaystyle\frac{t}{M_\pi^2} +   \lambda_{+,0}''   
\displaystyle\frac{t^2}{M_\pi^4}  
+ \dots \right)~ , 
\end{eqnarray}
where $t=(p_K - p_\pi)^2$.

As shown explicitly in Eq.~(\ref{eq:masterkl3}), it is convenient 
to normalise the form factors of all channels to $f_{+}^{K^0 \pi^-} (0)$, 
which in the following will simply be denoted by $f_{+}(0)$.
The channel-dependent terms 
$\Delta^K_{SU(2)}$ and $\Delta^{K \ell}_{\rm EM}$ represent
the isospin-breaking and long-distance electromagnetic corrections,
respectively. A determination of $V_{us}$ from $K_{\ell 3}$ decays at
the $1\%$ level requires $\sim 1 \%$ theoretical control 
on  $ f_{+}(0)$ as well as the inclusion of 
$\Delta^K_{SU(2)}$ and $ \Delta^{K \ell}_{\rm EM}$.

\subsubsection{SU(2) breaking and radiative corrections}

The natural framework to analyze these corrections is provided by
chiral perturbation
theory~\cite{Weinberg:1978kz,Gasser:1983yg,Gasser:1984gg} (CHPT), the
low energy effective theory of QCD.  Physical amplitudes are
systematically expanded in powers of external momenta of
pseudo-Goldstone bosons ($\pi, K , \eta$) and quark masses.  When
including electromagnetic corrections, the power counting is in
$(e^2)^{m} \, (p^2/\Lambda_\chi^2)^{n}$, with $\Lambda_\chi \sim 4 \pi
F_\pi$ and $p^2 \sim O(p_{\rm ext}^2 , M_{K,\pi}^2) \sim O(m_q) $.  To
a given order in the above expansion, the effective theory contains a
number of low energy couplings (LECs) unconstrained by symmetry alone.
In lowest orders one retains model-independent predictive power, as
these couplings can be determined by fitting a subset of available
observables. Even in higher orders the effective theory framework
proves very useful, as it allows one to bound unknown or neglected 
terms via power counting and dimensional analysis arguments.

Strong isospin breaking effects $O(m_u - m_d)$ were first studied to
$O(p^4)$ in Ref.~\cite{Gasser:1984ux}.  Both loop and LECs contributions
appear to this order. Using updated input on quark masses and the
relevant LECs, the results quoted in Table~\ref{tab:radcorr} for
$\Delta^K_{SU(2)}$ were obtained in Ref.~\cite{radcorr1}.

Long distance electromagnetic corrections were studied within CHPT to
order $e^2 p^2$ in Refs.~\cite{radcorr1,radcorr1bis}.  To this order,
both virtual and real photon corrections contribute to $\Delta^{K
\ell}_{\rm EM}$.  The virtual photon corrections involve (known) loops
and tree level diagrams with insertion of $O(e^2 p^2)$ LECs.  Some of
these LECs have been estimated in~\cite{moussallam}, using large-$N_C$
techniques. The remaining LECs have been bounded by dimensional
analysis (although they could be in principle estimated with the same
techniques as in~\cite{moussallam}). The resulting uncertainty is
reported in Table~\ref{tab:radcorr}, and does not affect the
extraction of $V_{us}$ at an appreciable level.

Radiation of real photons is also an important ingredient in the
calculation of $\Delta^{K \ell}_{\rm EM}$, because only the inclusive
sum of $K_{\ell3}$ and $K_{\ell 3 \gamma}$ rates is infrared finite to
any order in $\alpha$. Moreover, the correction factor depends on the
precise definition of inclusive rate.  In Table~\ref{tab:radcorr} we
collect results for the fully inclusive rate (``full'') and for the
``3-body'' prescription, where only radiative events consistent with three-body
kinematics are kept. CHPT power counting implies that to order
$e^2 p^2$ one has to treat $K$ and $\pi$ as point-like (and with
constant weak form factors) in the calculation of the radiative rate,
while structure dependent effects enter to higher order in the
chiral expansion ~\cite{kl3rad}.

Radiative corrections to $K_{\ell 3}$ decays
have been recently calculated also outside the CHPT
framework~\cite{radcorr2,radcorr3}.  Within these schemes, the UV
divergences of loops are regulated with a cutoff (chosen to be around
1 GeV).  In addition, the treatment of radiative decays includes part
of the structure dependent effects, introduced by the use of form
factors in the weak vertices.  Table~\ref{tab:radcorr} shows that
numerically the ``model'' approach of Ref.~\cite{radcorr2} agrees
rather well with the effective theory.

Finally, it is worth stressing that the 
consistency of the calculated strong and SU(2) corrections can be 
probed by experimental data by comparing the determination of
the universal term $V_{us} \times f_{+} (0)$ from the various decay modes,
as we will discuss in section~\ref{sect:Vus_kl3}.

\begin{table}[t]
\centering
\begin{tabular}{r||r||r|r}
&  $\Delta^K_{SU(2)}$ (\%) 
& \multicolumn{2}{|c}{
 $\Delta^{K \ell}_{\rm EM}(\%)$}     \\
&  &
\multicolumn{2}{|c}{ 3-body $\qquad\qquad$ full}  \\
\hline
$K^{+}_{e3}$ & 2.31 $\pm$ 0.22 ~ \cite{Gasser:1984ux,radcorr1}
&  -0.35  $\pm$ 0.16 ~ \cite{radcorr1}
& -0.10 $\pm$ 0.16 ~ \cite{radcorr1} \\
\hline 
$K^{0}_{e 3}$ &  0     &   +0.30  $\pm$  0.10 ~ \cite{radcorr1bis}
&   +0.55 $\pm$ 0.10   ~ \cite{radcorr1bis}  \\
              &             &                      & 
+0.65 $\pm$ 0.15  ~ \cite{radcorr2} \\
\hline 
$K^{+}_{\mu 3}$ & 2.31 $\pm$ 0.22 ~ \cite{Gasser:1984ux,radcorr1}
 &                    &              \\
\hline 
$K^{0}_{\mu 3}$ &  0               &               & 
+0.95 $\pm$ 0.15  ~ \cite{radcorr2} \\
\end{tabular}
\caption{
Summary of SU(2) and radiative correction factors for various $K_{\ell 3}$ 
decay modes.  Refs.~\cite{Gasser:1984ux,radcorr1,radcorr1bis} work within 
chiral perturbation theory to order $p^4, e^2 p^2$, while 
Ref.~\cite{radcorr2} works within a hadronic model for Kaon electromagnetic 
interactions.  
}
\label{tab:radcorr}
\end{table}

\subsubsection{Analytic results on  $f_{+}(0)$}

Within CHPT we can  
break up the form factor according to its expansion in quark masses:
\begin{equation}
f_{+}(0) = 1 + f_{p^4} + f_{p^6} + \dots \ . 
\end{equation}
Deviations from unity (the octet symmetry limit) are of second order
in SU(3) breaking~\cite{ag}.  The first correction arises to
$O(p^4)$ in CHPT: a finite one-loop
contribution~\cite{Gasser:1984ux,Leutwyler:1984je} determines
$f_{p^4}= - 0.0227$ in terms of $F_\pi$, $M_K$ and $M_\pi$, with
essentially no uncertainty.  The $p^6$ term receives contributions
from pure two-loop diagrams, one-loop diagrams with insertion of one
vertex from the $p^4$ effective Lagrangian, and pure tree-level
diagrams with two insertions from the $p^4$ Lagrangian or one
insertion from the $p^6$ Lagrangian~\cite{Post:2001si,Bijnens:2003uy}:
\begin{equation}
f_{p^6} = f_{p^6}^{\rm 2-loops} (\mu) +   
f_{p^6}^{L_i \times {\rm loop}} (\mu) 
+  f_{p^6}^{\rm tree} (\mu)    \ . 
\end{equation}
Individual components depend on the chiral renormalization scale
$\mu$, their sum being scale independent. 
Using as reference scale $\mu=M_\rho=0.77$ GeV and the $L_i$ from fit
10 in Ref.~\cite{Amoros:2001cp}, one has~\cite{Bijnens:2003uy}: 
\begin{equation}
f_{p^6}^{\rm 2-loops} (M_\rho) =  0.0113  \ , \qquad  
f_{p^6}^{L_i \times {\rm loop}} (M_\rho) = - 0.0020 \pm 0.0005 \ .
\label{eq:Liloop}     
\end{equation}
The explicit form for the tree-level contribution in terms of the LECs
$L_5$~\cite{Gasser:1984gg} and $C_{12,34}$~\cite{Bijnens:1999sh} is
then~\cite{Bijnens:1998yu,Bijnens:2003uy}
\begin{equation}
\label{eq:tree1}
f_{p^6}^{\rm tree} (M_\rho) = 
8 \frac{\left( M_K^2 - M_\pi^2 \right)^2}{F_\pi^2}  
\, \left[\frac{\left(L_5^r (M_\rho) \right)^2}{F_\pi^2} - 
C_{12}^r (M_\rho) - C_{34}^r (M_\rho) \right] \ .  
\end{equation}
$L_{5}^r (M_\rho)$ is known from phenomenology to a level that 
induces less than $1\%$ uncertainty in $f_{p^6}^{\rm tree} (M_\rho)$.
The $p^6$ constants $C_{12,34}$ can in principle be determined
phenomenologically.  It has been shown in Ref.~\cite{Bijnens:2003uy}
that combinations of $C_{12}$ and $C_{34}$ govern the slope
$\lambda_0$ and curvature $\lambda_0''$ of the scalar form factor
$f_{0}(t)$, accessible in $K_{\mu 3}$ decays.  In order to extract
$C_{12}$ and $C_{34}$ to a useful level (i.e. leading to $1 \%$ final
uncertainty in $V_{us}$), one needs experimental errors at the level
$\Delta \lambda_0 \sim 0.001 $ (roughly a $5 \%$ measurement) and
$\Delta \lambda_0'' \sim 0.0001 $ (roughly a $20 \%$ measurement), as
well as $F_K/F_\pi$ at the $1 \%$ level from theory.

\subsubsection*{Large-$N_C$ estimate of $f_{p^6}^{\rm tree}$}

In Ref.~\cite{spp05} a (truncated) large-$N_C$ estimate of
$f_{p^6}^{\rm tree}$ was performed.  It was based on matching a
meromorphic approximation to the $\langle S P P \rangle$ Green
function (with poles corresponding to the lowest-lying scalar and
pseudoscalar resonances) onto QCD by imposing the correct 
large-momentum falloff, both off-shell and on one- and two-pion mass
shells. 
In particular, $C_{12}$ is uniquely determined by requiring the
correct behavior of the pion scalar form factor $ \langle \pi | S |
\pi \rangle $, while $C_{34}$ is fixed by the correct scaling of the
one-pion form factors $ \langle \pi | S | {\cal P} \rangle $ and $\langle
\pi | P | {\cal S} \rangle$.
The uncertainty of the large-$N_C$ matching procedure was estimated by
varying the chiral renormalization scale at which the matching is
performed in the range $\mu \in [M_\eta, 1 {\rm GeV}]$, and is found
to be $ \delta f_{p^6}^{\rm tree}|_{1/N_C} =  \pm 0.008$.  
The final result is 
\begin{equation}
\label{eq:tree2}
f_{p^6}^{\rm tree} (M_\rho) = 
-  \frac{\left( M_K^2 - M_\pi^2 \right)^2}{2 \, M_S^4}  
\, \left( 1 - \frac{M_S^2}{M_P^2} \right)^2  = 
- 0.002  \pm 0.008_{\, 1/N_C} \pm 
0.002_{\, M_S}   \ , 
\end{equation}
and is much smaller than the ratio of
mass scales $(M_K^2 - M_\pi^2)^2/M_S^4 $ would suggest, due to
interfering contributions. When combined with the $p^6$ loop 
corrections \cite{Bijnens:2003uy}, this estimate leads 
to $f_{p^6} = 0.007 \pm 0.012$.
Variations of the hadronic ansatz lead to the conclusion that the
smallness of the tree-level part compared to the loop contribution of
$O(p^6)$ for $f_{+}(0)$ appears as generic feature of a
few-resonance approximation for the set of large-momentum constraints
considered. As a consitency check of this approach, it is 
worth to mention that within the same framework  
one obtains a prediction for the slope of the scalar form factor, 
$ \lambda_0 = 0.013 \pm 0.002_{\,
1/N_C} \pm 0.001_{\, M_S} $, fully consistent with the value measured
by KTeV in $K^L_{\mu3}$ decays, $\lambda_0 =
(13.72 \pm 1.31) \times 10^{-3}$ \cite{Alexopoulos:2004sy}.

Combining this $p^6$ result with the well-know  $p^4$ term,
leads to the  following global estimate of $f_{+}(0)$:
\begin{equation}
f_{+}(0)_{{\rm large}-N_C} = 0.984 \pm 0.012~.
\end{equation}
This value is substantially higher --although
compatible within the errors--  with respect 
to the old estimate 
by Leutwyler Roos \cite{Leutwyler:1984je}
\begin{equation}
f_{+}(0)_{\rm Leutwyler-Roos} = 0.961 \pm 0.008~,
\label{eq:LRoss}
\end{equation}
which for a long time has been the reference 
value of $f_{+}(0)$ (and it is still the value 
adopted by the PDG \cite{PDG2004}) 
in the extraction of $V_{us}$.

\subsubsection{$ f_{+}(0)$ from  lattice QCD}
Starting from Ref.~\cite{becirevic},
in the last few months it has been realized that lattice QCD
is an excellent tool to estimate $f_{+}(0)$ at a level of 
accuracy interesting for phenomenological purposes
\cite{FNAL,Okamoto2,Okamoto1,JLQCD,RBC}.

On general grounds, determining a form factor at the $1\%$ level 
of accuracy seems very challenging --if not impossible--
for present lattice-QCD calculations. However, the specific case
of $f_{+}(0)$ is quite special:
by appropriate ratios of correlation functions 
one can directly  isolate the SU(3)-breaking quantity $[f_{+}(0)-1]$, 
or even better the quantity $[f_{+}(0)-1-f_{p^4}]$, which is the 
only irreducible source of uncertainty~\cite{becirevic}.
Estimating these SU(3)-breaking quantities with a relative error of about $30\%$ 
is sufficient to predict $f_{+}(0)$ at the $1\%$ level or below. Thus even with the 
present techniques there are good prospects to obtain lattice 
estimates of $f_{+}(0)$ of phenomenological interest.

\begin{figure}[t]
\begin{center}
\hskip -1 cm
\epsfxsize=8.5 cm 
\epsfxsize=8 cm 
\hskip 0.5 cm
\epsffile{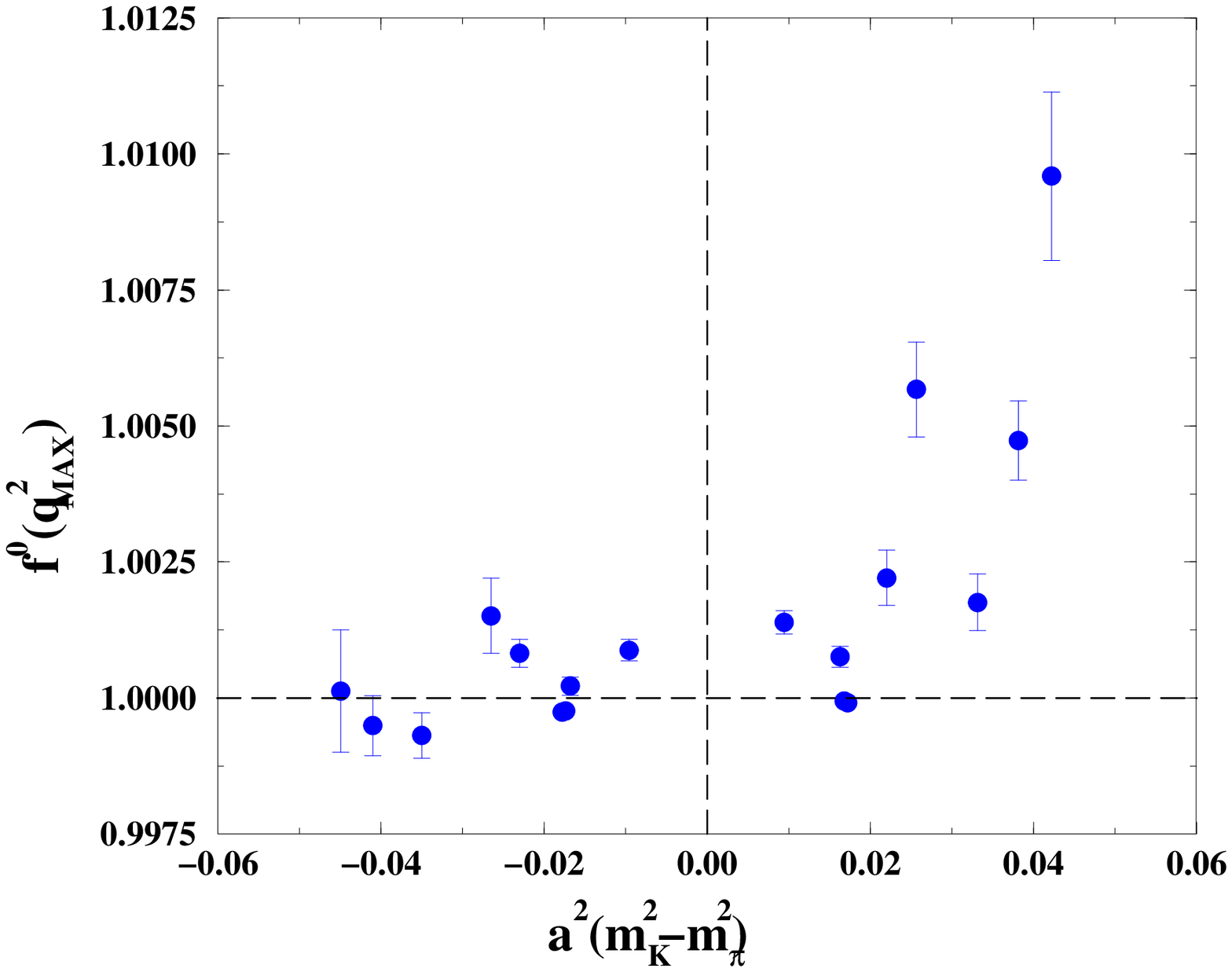}
\epsfxsize=7.5 cm 
\epsfysize=7 cm 
\epsffile{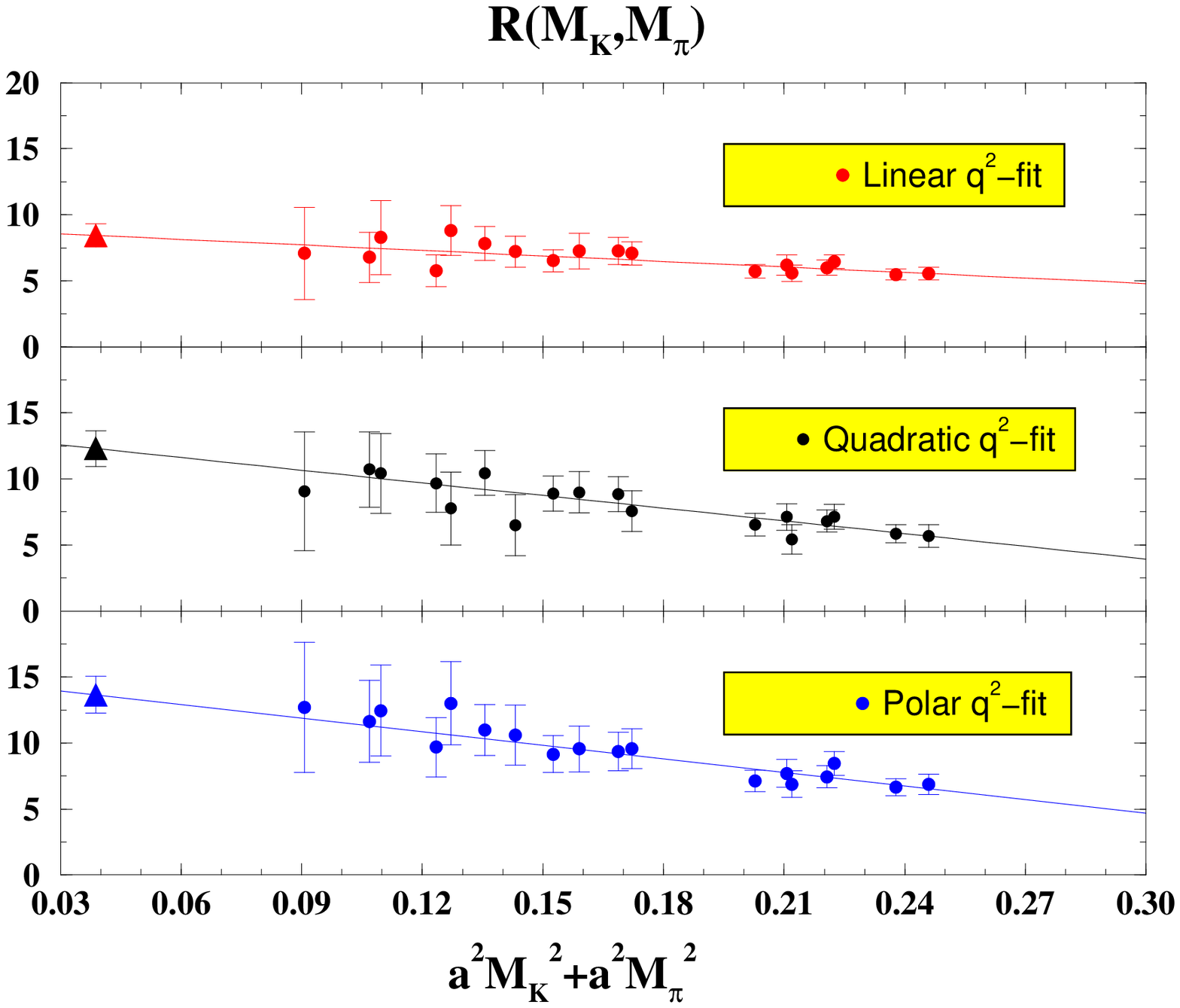}
\end{center}
\caption{Left: $f_0(q^2_{\rm max})$ as a function of $[M_K^2 - M_{\pi}^2]$ 
in units of the lattice spacing ($a$). Right:
$R(M_K, M_\pi)$ as a function of $[M_K^2 + M_{\pi}^2]$ 
for the cases of the linear, quadratic and polar $q^2$-extrapolation 
in the fit of $f_0(q^2)$; the triangles indicate the values 
of $R(M_K, M_\pi)$ extrapolated to the
physical kaon and pion masses.}
\label{fig:linfit} 
\end{figure}

The analysis of Ref.~\cite{becirevic} is based on the following 
three main steps: 
\begin{enumerate}
\item[I] {\em Evaluation of the scalar form factor $f_0(q^2)$ at 
      $q^2 = q^2_{\rm max}=(M_K-M_\pi)^2$}. 
Applying a method originally proposed in Ref.~\cite{FNAL} to investigate
heavy-light form factors, the scalar form factor is extracted from 
the following double ratio of matrix elements:
\beq
\frac{ \langle \pi | \bar{s} \gamma_0 u  | K \rangle \langle K | \bar{u} \gamma_0 s | \pi \rangle }{ 
       \langle \pi | \bar{u} \gamma_0 u | \pi \rangle \langle K | \bar{s} \gamma_0 s | K \rangle } =  
\frac{(M_K + M_\pi)^2}{4 M_K M_\pi}\left[ f_0(q^2_{\rm max}; M_K, M_\pi)
\right]^2 ~ ,
\label{eq:fnal}
\eeq 
where all mesons are at rest. The double ratio and the kinematical configuration 
allow to reduce most of the systematic uncertainties and to reach a statistical 
accuracy on $f_0(q_{\rm max}^2 )$ well below $1 \%$ (see figure~\ref{fig:linfit}).
 
\item[2] {\em Extrapolation of $f_0(q_{\rm max}^2)$ to $f_0(0)=f_+(0)$}. 
By evaluating the $q^2$ dependence of the form factor, 
the latter is extrapolated from $q_{\rm max}^2$ to $q^2 = 0$.  
This procedure is performed independently for various sets 
of meson masses (with corresponding light-quark masses chosen in the 
range $0.5 m_s$-- $2 m_s$), and using different functional
forms (linear, quadratic and polar) for the $q^2$ dependence.
A byproduct of this step is an estimate of the physical slope
of the scalar form factor: the prediction quoted in 
Ref.~\cite{becirevic} for this observable has recently been 
confirmed by experimental analysis 
of KTeV \cite{Alexopoulos:2004sy}.

\item[3] {\em Subtraction of the chiral logs and chiral extrapolation}.
The $f_+(0)$-values thus obtained needs to be extrapolated to
the physical values of $M_K$ and $M_\pi$.
In order to reduce the error of this extrapolation, and correct for 
the leading quenched artifacts, the following ratio is considered:
\beq 
R(M_K, M_\pi)  = \dfrac{\Delta f}{(\Delta M^2)^2}  \equiv 
 \dfrac{  1 + f^q_{p^4}(M_K, M_\pi) - f_+(0; M_K, M_\pi) }{ (M_K^2 - M_{\pi}^2)^2}~.
\label{eq:ratioR}
\ee
Here $f^q_{p^4}(M_K, M_\pi)$ denotes the $\cO(p^4)$
contribution evaluated within quenched CHPT which, similarly to 
its unquenched analog ($f_{p^4}$),  is finite and free from unknown 
counterterms. By construction, the ratio (\ref{eq:ratioR}) 
is finite in the SU(3) limit, does not depend on any subtraction 
scale, and is free from the dominant quenched chiral 
logs of  $\cO(p^4)$. Extrapolating $R(M_K, M_\pi)$ to the physical masses 
(see figure~\ref{fig:linfit}) leads to
$\Delta f = R(M_K^{\rm phys}, M_\pi^{\rm phys}) \times [(\Delta M^2)^2]^{\rm phys} = 
(0.017 \pm 0.005_{\rm stat} \pm 0.007_{\rm syst} )$,
which implies~\cite{becirevic}:
\be
f_{+}(0)_{\rm Lattice-quenched} = 0.960 \pm 0.005_{\rm stat} \pm 0.007_{\rm syst}~,
\label{eq:DF}
\ee
in remarkable agreement with the old estimate 
by Leutwyler and Roos [see Eq.~(\ref{eq:LRoss})].
\end{enumerate}

\noindent 
The result of Ref.~\cite{becirevic} is affected by an irreducible 
systematic error due to the residual effects [$\cO(p^6)$ and above]
of the quenched approximation. This error is difficult to quantify and is not included 
in the systematic error of Eq.~(\ref{eq:DF}), which is dominated by 
the extrapolation to the physical meson masses. However, the good
agreement between the slope of the scalar form factor obtained 
within this simulation and the experimental one 
seems to suggest that this additional source of uncertainty is 
not large. 

Very recently new analyses of $f_+(0)$ with unquenched lattice 
simulations ($N_f=2$ and  $N_f=2+1$) have been started by 
various groups \cite{Okamoto2}. In particular, the preliminary results by 
the collaborations 
MILC and HPQCD ($f_+(0)=0.962(6)(9)$  \cite{Okamoto1}), JLQCD
($f_+(0)=0.952(6)$ \cite{JLQCD}), and RBC  ($f_+(0)=0.955(12)$ \cite{RBC}) 
are very encouraging: the errors do not exceed the $1\%$ level
and the results are well compatible. Interestingly, these preliminary 
unquenched results for  $f_+(0)$ are also in good agreement 
with the quenched estimate of Ref.~\cite{becirevic}.

\subsection{The extraction of $V_{us}$ from $K_{\ell3}$ decays}
\label{sect:Vus_kl3}
In the last few years several experiments,
based on different techniques, have put a lot of effort in re-measuring 
all the observables of the main $K$ decay modes, whose 
values available in literature dated back to the 70's.  
As discussed above, the master formula for the 
extraction of $|V_{us}|$ from  
$K_{\ell 3}$ decays is Eq.~(\ref{eq:masterkl3}).
The experimental inputs are
the semileptonic widths (based on the semileptonic
branching fractions and lifetimes) and the form factors,
which  are necessary for the calculation
of the phase-space integrals.
Since none of the experiments have measured yet all of the experimental
inputs required to calculate $|V_{us}|$ independently, 
we have calculated average values of
$|V_{us}|f_+(0)$ for $K^+(e3)$, $K_L(e3)$, $K_L(\mu3)$, and $K_S(e3)$ using 
the inputs described in the following sections.  
We then combine these results to find an average $|V_{us}|f_+(0)$ 
for $K_{\ell 3}$ decays. 

\subsubsection{Charged and neutral kaon branching fractions}
Three experiments have contributed to the measurement of the 
neutral kaon branching fractions: KTeV, NA48 and KLOE. The first two are fixed target 
detectors using secondary high energy neutral beams and situated downstream
from the primary target to obtain a beam of $K_{L}$, with small
contamination of $K_{S}$. No absolute kaon count is available so that one
can measure ratios of partial widths. 
Instead KLOE is a collider experiment at a $\phi$-factory where $K_{S}-K_{L}$
pairs are produced and therefore takes advantage of the 
unique feature of the {\it tagging}:
identified $K_L(K_S)$ decays tag a $K_S(K_L)$ beam and
provide the means for measurements of absolute branching ratios.
Therefore the measurements performed by the three experiments have
quite different systematic uncertainties.

\begin{figure}[t]
\centering
\psfig{figure=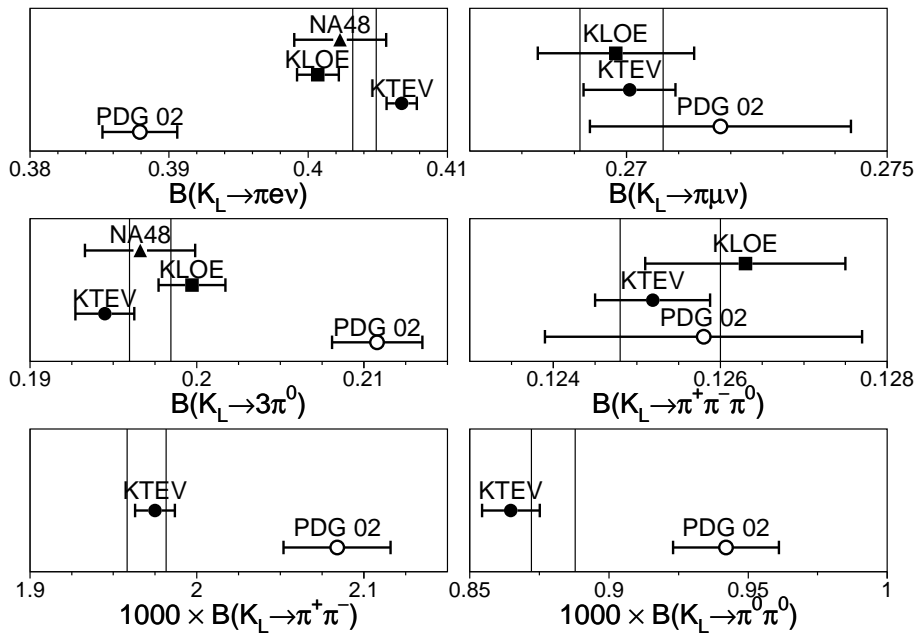,width=12 cm}
\caption{ 
    \label{fi:pdg2} 
       $K_L$ branching fractions measured  by KTeV, KLOE, NA48, and from the 
       PDG 2002
       fit (open circles). The vertical lines indicate the $\pm 1\sigma$ bounds
       from a fit to all KTeV, KLOE, and NA48 measurements.
       }
\end{figure}

For the $\KL$ branching fractions, we consider the following 
experimental inputs:
\begin{itemize}
\item KTeV measured the following 5 partial width ratios~\cite{ktev_kbr}: 

$\Rpimunu/ \Rpienu$~, $\Rppp/ \Rpienu,$  \\
$\Rzzz/ \Rpienu$~, $\Rpm/ \Rpienu$~, \\
$\Rzz/ \Rzzz$. 

Since the six decay modes listed above account for more than
99.9\% of the total decay rate, the five partial width ratios may be 
converted into measurements of
the branching fractions for the six decay modes. 

\item KLOE uses a tagged $K_L$ sample to measure the 4 largest $K_L$ branching 
fractions~\cite{kloe_kbr}.

\item NA48 measures the following 2 ratios~\cite{na48_ke3}: 

$\Rpienu/ \Gamma(\KL \to 2~{\rm tracks})$, $\Rzzz/\Gamma(\KS \to \pi^0\pi^0)$,   \\
which can be used to determine $B(K_{e3})$ and $B(3\pi^0)$.

\end{itemize}
A critical issue is represented by the inclusiveness of the measurement and 
the treatment of the radiative photon. These have been carefully accounted
for in all the experiments. 
%
%
A fit to all of these measurements, accounting for correlations, gives the
$K_L$ branching ratios summarized in Table~\ref{ta:br}.  
Figure~\ref{fi:pdg2} shows a comparison of 
these measurements along with the best fit values for each of the six 
branching fractions.

\begin{table}[t]
\centering
\begin{tabular}{ lcc}
\hline
\hline
Decay Mode       & Branching fraction & $\Gamma_i$ ($10^7 s^{-1}$) \\
\hline 

\KLpienu & $0.4040 \pm 0.0008$ & $0.7908 \pm 0.0032$ \\
\KLpimunu & $ 0.2699 \pm 0.0008 $& $0.5283 \pm 0.0023$  \\
\KLpmz & $ 0.1253 \pm 0.0006 $ &  $0.2452 \pm 0.0015$\\
\KLzzz & $ 0.1972 \pm 0.0012  $ & $0.3859 \pm 0.0029$  \\
\KLpm & $(1.971 \pm 0.012 )\times 10^{-3}  $ 
       &  $(3.857 \pm 0.027 )\times 10^{-3}  $  \\
\KLzz & $(0.880 \pm 0.008)\times 10^{-3}  $ 
& $( 1.722 \pm 0.017 )\times 10^{-3} $ \\
\hline \hline 
\end{tabular}
\caption{ 
    \label{ta:br}
     Average $K_L$ branching fractions and widths 
     based on fit to all new measurements
     from KTeV, KLOE, and NA48. The width measurements use the PDG 2004
     average $K_L$ lifetime combined with the two preliminary KLOE measurements
     mentioned in Section~\ref{sec:life}.
       }
\end{table}

For $K_S$, we use the KLOE preliminary measurement:
$B(\KS \to \pi e \nu) = (7.09 \pm 0.09) \times 10^{-4}$~\cite{kloe_kse3}.
For $K^\pm$, we average the BNL E865~\cite{bnl865} 
measurement, $B(K^\pm \to \pi^0 e^{\pm}\nu)=(5.13 \pm 0.10)\%$,   
with the more recent NA48 result,  
$B(K^\pm \to \pi^0 e^{\pm}\nu)=(5.14 \pm 0.06)\% $, finding 
$B(K^\pm \to \pi^0 e^{\pm}\nu)=(5.14 \pm 0.06)\% $.

\subsubsection{Lifetime Measurements}
\label{sec:life}
KLOE presented two new measurements of the $\KL$ lifetime: an ``indirect method,''
 based on the $K_L$ lifetime required to make $\Sigma B_i =1$, exploiting the lifetime 
dependence of the detector acceptance~\cite{kloe_kbr}, and a ``direct method'', based on 
the fit of the proper time distribution of $\KL \to 3\pi^0$ decays~\cite{taukl_kloe}. 
These new results and the old PDG average are listed in 
Table~\ref{ta:life}. The new average value, which we use for the results
quoted below, is $\tau_L = (50.98 \pm 0.21)$ ns.

Combining the $K_L$ branching 
fractions with the new lifetime gives the partial decay
widths quoted 
in Table~\ref{ta:br}. Note that correlations between the KLOE branching
fractions and the ``indirect'' KLOE lifetime determination have been included.
For the $\KS$ and $\KP$ lifetimes, we use the PDG averages. 

It is worthwhile to note that KLOE is about to report new results 
on the $\K^\pm$ lifetimes and the achievable accuracy is at the few per mil level. 
The measurement is done
with two independent methods: the first one is based on the charged kaon
decay path while the second one measures directly the kaon time of flight
using the photons from $\pi^{0}\to\gamma\gamma$ decay, using the charged kaon
decay channels with a $\pi^{0}$ in the final state. This will allow the
clarification of the present situation which shows discrepancies between
``in-flight" and ``at-rest" charged kaon lifetime measurements, used in the
PDG average.  
\begin{table}[t]
\centering
\begin{tabular}{ lc}
\hline
\hline
Source       & Lifetime (ns)\\
\hline
PDG 2004 Average & $51.5 \pm 0.4$ \\
KLOE (``indirect'') & $50.72 \pm 0.35$ \\
KLOE (``direct'') & $ 50.87 \pm 0.31$\\
\hline
New Average & $50.98 \pm 0.21$ \\
\hline \hline 
\end{tabular}
\caption{ 
    \label{ta:life}
     $K_L$ lifetime measurements.
       }
\end{table}
\subsubsection{Phase Space Integrals}
Recent experiments have also performed new measurements of the semileptonic form factors needed
to calculate the phase space integrals:
\begin{itemize}
\item KTeV has measured form factors in both $K^0_{e3}$ and $K^0_{\mu3}$ \cite{ktev_kl3ff}.
\item NA48 has measured the $K^0_{e3}$ form factor~\cite{na48_ff}.
\item ISTRA+ has measured the $K^-_{e3}$ form factor~\cite{istra_ff}.
\end{itemize}
KLOE is also measuring the form factors in both neutral and charged kaon decays, 
although final results have not been published yet. Note that, in principle, 
KLOE is the only experiment with the possibility of measuring all
the useful inputs for the extraction of $V_{us}$, namely lifetimes,
branching fractions and form factors both for neutral and charged kaons.

In the present analysis, 
to calculate phase space integrals, we use the KTeV quadratic form factor results
for neutral kaon decays and the
ISTRA+ quadratic form factor measurements for charged kaons.  For both
charged and neutral decays, we include an additional 0.7\%\ uncertainty to
the phase 
space integrals, as suggested by KTeV~\cite{ktev_kl3ff}, to account for
differences between the quadratic and pole model 
form factor parameterizations, both of which give acceptable fits to the
data. The resulting phase space integrals are 
$\textstyle I^{K^0 e} = 0.1535 \pm 0.0011  $, 
$\textstyle I^{K^0 \mu} = 0.10165 \pm 0.0008 $,
and  $\textstyle I^{K^+ e} = 0.1591 \pm 0.00012$.
\subsubsection{$K_{\ell 3}$ results for $|V_{us}|$}
We are now ready to extract four estimates of $f_+(0)|V_{us}|$ 
from Eq.~(\ref{eq:masterkl3}), using the available data on 
 $\KL$ (both $e$ and $\mu$ modes), $K^{\pm}$ and $\KS$ decays.
To this purpose, we set $S_{EW}=1.023$ ~\cite{sew}
and use the SU(2) and radiative correction factors 
reported in Table~\ref{tab:radcorr}.

\begin{figure}[t]
\centering
\begin{center}
\includegraphics[width=10.5 cm,angle=-90]{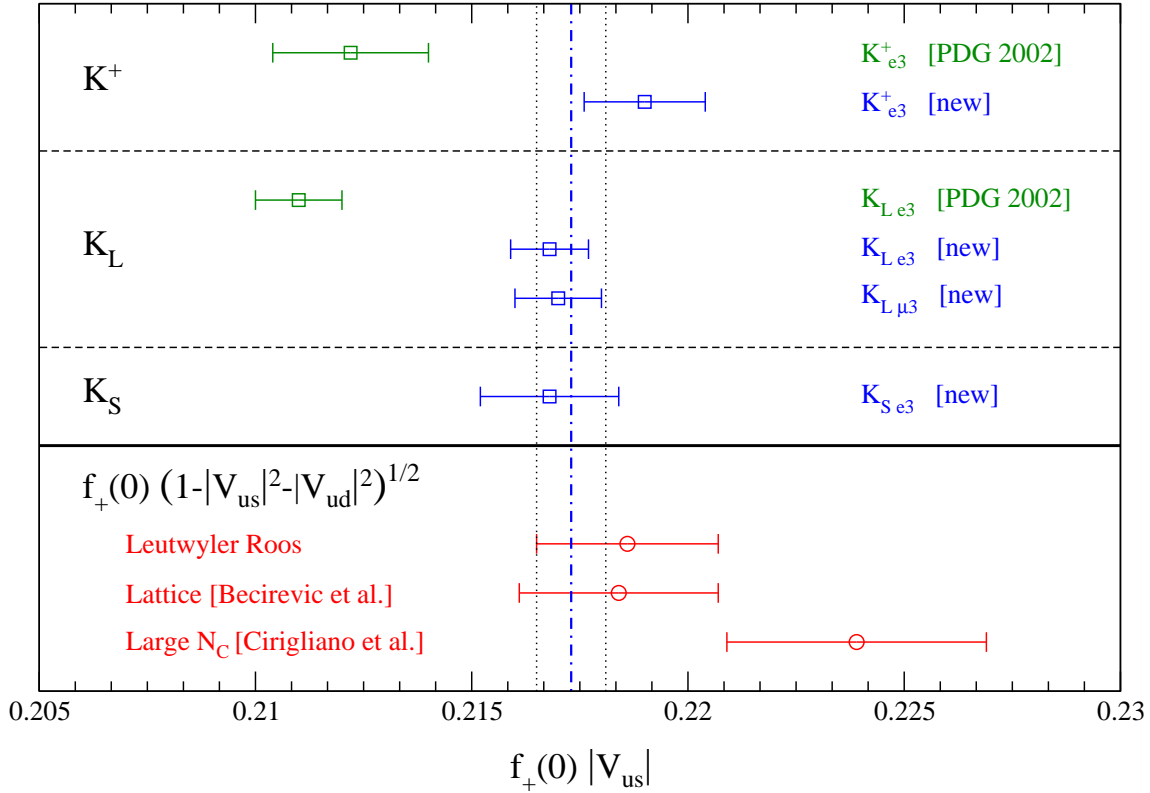}
\end{center}
\caption{
Comparison of the recent measurements of $|V_{us}|f_+(0)$ 
with the PDG 2002 results, and with the expectations 
of unitarity based on different theoretical calculations of 
$f_+(0)$~\cite{Leutwyler:1984je,spp05,becirevic}. 
The values of $f_+(0)(1-|V_{ud}|^2-|V_{ub}|^2)^{1/2}$ have been obtained 
using the 2005 value of $V_{ud}$ in Eq.~(\ref{eq:Vud2005});
the error bars are largely dominated by the theoretical 
uncertainty on $f_+(0)$. The vertical lines correpsond to 
the average value of $f_+(0)|V_{us}|$ in Eq.~(\ref{eq:fin_kl3}).}
\label{fig:vus}
\end{figure}

Figure~\ref{fig:vus}  shows a comparison of the PDG and the 
averages of recent measurements for $f_+(0)|V_{us}|$ for the various 
decay modes. The figure also shows 
$f_+(0)(1-|V_{ud}|^2-|V_{ub}|^2)^{1/2}$,
namely the expectation for $f_+(0)|V_{us}|$ assuming unitarity,
based on $|V_{ud}| = 0.9738 \pm 0.0003$, 
$|V_{ub}| = (3.6 \pm 0.7)\times 10^{-3}$,
and several recent calculations of $f_+(0)$.
The average of all recent measurements gives
\begin{equation}
f_+(0)|V_{us}| = 0.2173 \pm 0.0008
\label{eq:fin_kl3}
\end{equation}
Using the Leutwyler-Roos estimate of $f_+(0)$,
this implies
\be
 |V_{us}|_{K_{\ell 3}} = 0.2261 \pm 0.0021  
  \qquad  \left[ f_+(0) = 0.961 \pm 0.008 \right]~.
\ee

%
%
\subsection{$V_{us}$ from $K_{\mu 2}$}
As recently pointed out in Ref.~\cite{Marciano},
besides the traditional method of extracting $|V_{us}|$ 
from $K_{\ell 3}$ decays, one can obtain an independent and competitive 
estimate of $|V_{us}|$ (or, to be more precise, of $|V_{us}/V_{ud}|$)
from the ratio of the inclusive decay widths 
of $\kmuii$ and $\pi^{+}\rightarrow\mu^{+}\nu(\gamma)$ 
decays.  The measurable ratio can be written as follows 
\begin{equation}
\frac{\Gamma(K^+ \to \mu^+\nu(\gamma))}{\Gamma(\pi^+ \to \mu^+\nu(\gamma))}=\frac{V_{us}^2}
{V_{ud}^2}\frac{F_K^2}{F_\pi^2}\frac{M_\pi^3 (M_K^2-M_\mu^2)^2 }{M_K^3 (M_\pi^2-M_\mu^2)^2}\left[
1-\frac{\alpha}{\pi}(C_\pi-C_K)\right]~,
\label{eq:Km2}
\end{equation}
where $F_{K,\pi}$ denote kaon and pion decay
constants, and $C_{K,\pi}$ parametrize the radiative-inclusive electroweak
corrections, accounting for soft-photon emission and
 corresponding virtual corrections. 
According to the detailed analyses of Refs.~\cite{Knecht, Cirigliano_priv}:
$C_\pi - C_K = 3.0 \pm 0.75$. 

The key theoretical input for this method is the estimate of $F_K/F_\pi$,
obtained by means of lattice QCD. Contrary to the case of $f_+(0)$, 
which is protected by the Ademollo--Gatto theorem, 
the quantity $[F_K/F_\pi-1]$ breaks SU(3) invariance already at 
the first order. It is therefore very challenging to estimate 
$F_K/F_\pi$ at the $1\%$ level of accuracy. However, the MILC 
collaboration has recently shown that this is possible \cite{MILC}.
In 2004 they found $F_K/F_\pi=1.210 \pm 0.014$ \cite{MILC}, value
which is compatible with the preliminary 
result of their new analysis, namely
$F_K/F_\pi=1.198 \pm 0.003^{+ 0.016}_{ - 0.005}$ \cite{MILC2}.

On the experimental side, the dominant error in Eq.~(\ref{eq:Km2})
is induced by the kaon decay width.
KLOE has recently performed a new measurement of the corresponding 
absolute branching ratio, obtaining 
$B(K^+ \to \mu^+ \nu (\gamma))=0.6366 \pm 0.009 \pm 0.00015$~\cite{kloe_kmu2}. 
This measurement is fully inclusive of the final
state radiation and is based on a sample of $K^{+}$
events tagged by $K^{-}$ decays produced at the 
$\phi$-meson peak, thus avoiding normalization issues such as trigger and
reconstruction efficiency (which enter at first order in the determination of
the branching ratio). 

Using the new KLOE result, the $K^+$ life time from PDG
and $V_{ud}=0.9738\pm0.0003$ (see section~\ref{sect:Vud_SFT}), 
leads to 
\bea
|V_{us}|_{\kmuii} &=& 0.2223 \pm 0.0026 \qquad (F_K/F_\pi=1.210 \pm 0.014)~, \\
|V_{us}|_{\kmuii} &=&  0.2245^{+0.0011}_{-0.0031} \qquad\ \ \
\left(F_K/F_\pi =1.198 \pm 0.003^{+0.016}_{-0.005} \right)~, 
\eea 
In both cases the error is largely dominated by the 
uncertainty on $F_K /F_{\pi}$ and, in particular, the uncertainty 
induced by $|V_{ud}|$ is negligible. Both results are consistent 
with the  $V_{us}$ determination from $K_{\ell 3}$.

\section{Alternative approaches to $V_{us}$}

\subsection{$V_{us}$ from hyperon decays: status and perspectives  }
Hyperon semileptonic decays (HSD), denoted here as $B_1 \to B_2 +\ell^-
+\overline{\nu}_\ell$, are described by the
$V-A$ theory. This theory states that the weak transitions arise from the
self-coupling of a single charged
current, which is the sum of the current operators for the leptons and the
strongly interacting particles. Each
current can be expressed as a linear combination of vector and axial
terms. The matrix elements of the lepton
current are unambiguous. In contrast, non-perturbative strong interaction
effects at low energy weak interactions
force the introduction of phenomenological form factors to account for
strong interaction effects in the matrix
elements. The transition amplitude for HSD can be written as \cite{gk}
\begin{equation}
{\cal M}_0 = \frac{G_F V_{CKM}}{\sqrt 2} \, [\overline u_\ell (l)
\gamma^\mu(1-\gamma_5) v_\nu (p_\nu)]
[ \overline u_{B_2} (p_2) W_\mu u_{B_1} (p_1)], \label{eq:m0}
\end{equation}
where $V_{CKM}$ is either $V_{ud}$ or $V_{us}$ and
\begin{eqnarray}
W_\mu = f_1 (q^2) \gamma_\mu + \frac{f_2 (q^2)}{M_1} \sigma_{\mu \nu}
q^\nu +  \frac{f_3 (q^2)}{M_1}q_\mu + \left[
g_1 (q^2) \gamma_\mu + \frac{g_2 (q^2)}{M_1}\sigma_{\mu \nu} q^\nu +
\frac{g_3 (q^2)}{M_1} q_\mu \right] \gamma_5.
\end{eqnarray}
Here $q \equiv p_1 - p_2$ is the momentum transfer and $M_1$ is the mass of
the decaying hyperon. The quantities $f_i(q^2)$ and $g_i(q^2)$ are the
conventional vector and axial-vector form factors, which are required to be
real by time reversal invariance.

The usual approach to evaluate the expected properties of the weak
hadronic current is based on the flavor SU(3) symmetry of the strong
interactions. In the limit of exact SU(3) symmetry, the hadronic weak
currents belong to SU(3) octets, so that the form factors of different HSD
are related to each other. The vector part of the weak current and the
electromagnetic current belong to the same octet. Thus, the vector form
factors are related at $q^2=0$ to the electric charge and the anomalous
magnetic moments of the nucleons. The conservation of the electromagnetic
current implies that $f_3(q^2)$ vanishes for all HSD in the SU(3) limit.
Similarly, $g_1$ is given in terms of two reduced form factors $F$ and $D$
whereas $g_2$ for diagonal matrix elements of hermitian currents vanishes by
hermiticity and time-reversal invariance. SU(3) symmetry then implies that
$g_2=0$ in the symmetry limit.

Currently, experiments in HSD allow precise measurements of form factors
(for a review about the experimental situation of HSD see for instance
Ref.~\cite{csw}). The statistical errors of these experiments are rather
small, and more effort has been put into the reduction of the systematic
errors, which can be of two types. The first one comes from the different
shortcomings of the experimental devices. The second one, of theoretical
nature, comprises i) radiative corrections and ii) theoretical assumptions
for some form factors, including their $q^2$ dependence. Let us briefly
discuss each one.

The analysis of low, medium, and high-statistics experiments of HSD
requires the inclusion of radiative corrections. Since no first principle
calculation of radiative corrections is yet possible, these corrections are
committed to model dependence and the experimental analyses which use them
become model-dependent too. Up to order $(\alpha/\pi)(q/M_1)$, the model 
dependence of radiative corrections can be absorbed into the form factors
originally defined in the matrix elements of the hadronic current
(\ref{eq:m0})~\cite{gk} and the remaining part is model independent. Within
these orders of approximation one is left with general expressions which can
be used in model-independent analyses \cite{rfm05}. Among the integrated
observables in HSD, only the decay rates need to be corrected whereas the
angular correlations and spin-asymmetry coefficients are practically
unaffected \cite{gk}.

On the other hand, the assumptions on the form factors are subtle to handle.
In particular, their $q^2$-dependence cannot always be neglected, since
noticeable contributions can arise. In order to obtain expressions correct
to order ${\mathcal O}(q^2)$, the $q^2$-dependence of $f_2$ and $g_2$ can be
ignored, because they already contribute to order ${\mathcal O}(q)$ to the
decay rate. For $f_1(q^2)$ and $g_1(q^2)$, instead, a linear expansion in
$q^2$ must be considered, because higher powers amount to negligible
contributions to the decay rate. A dipole parametrization of the leading
form factors works fine \cite{gk}.

Another important issue that must be taken into account is the validity of
the exact SU(3) limit. Presently, experiments are precise enough that exact
SU(3) no longer yields a reliable fit. For HSD, due to the presence of the
axial current, SU(3) breaking (SB) can occur in first order. This fact makes
HSD to be apparently less reliable to use for determining $V_{us}$ than
$K_{l3}$ decays, where the Ademollo-Gatto theorem reduces the effects of SB.
For this reason the present value of $V_{us}$ quoted by the Particle Data
Group \cite{PDG2004} is essentially derived from $K_{l3}$, while the one from
HSD is discarded.

Although there are various treatments on the calculation of SB in HSD
\cite{don,sch,kra,and,villadoro}, it is hard to assess their success 
because their predictions vary substantially from one another and none of
them can be considered as fully consistent. These calculations incorporate
second-order SB corrections into $f_1$. Some computations \cite{don,sch,kra}
find that $f_1$, for $|\Delta S|=1$ decays, is reduced from its symmetry
limit value $f_1^{\rm SU(3)}$ such that $f_1/f_1^{\rm SU(3)}<1$, whereas the
calculation of Ref.~\cite{and}, performed in the framework of Heavy Baryon
Chiral Perturbation Theory (HBChPT)~\cite{hbchpt}, found large and positive
corrections. Recently, the calculation of SB corrections within HBChPT
has been reconsidered in Ref.~\cite{villadoro}, where it is found that the
calculation of~\cite{and} contained some mistakes and missed some important
contributions. The result of Ref.~\cite{villadoro} is that SB corrections,
as obtained from HBChPT, are still positive but smaller than previously
claimed in~\cite{and}. In addition, the quantitative estimate of these
corrections strongly depends on the values used for the low energy constants
and does not include the contribution coming from the baryon decuplet. This
suggests that the corrections coming from higher orders can be large, and
that the present estimates may not be reliable. Alternatively, SB effects
can be also extracted from the data, for instance parameterising them in the
framework of the $1/N_c$ expansion of QCD \cite{djm95,dai,rfm98,rfm04}.

Fits to the experimental data of HSD to extract $V_{us}$ can be performed
using the decay rates and the spin and angular correlation coefficients
\cite{PDG2004}. There are sufficient data from five decays to make it
possible: $\Lambda \to p e^- \overline \nu_e$, $\Sigma^- \to n e^-\overline
\nu_e$, $\Xi^- \to \Lambda e^- \overline \nu_e$, $\Xi^-\to \Sigma^0 e^-
\overline \nu_e$, and $\Xi^0\to \Sigma^+ e^- \overline \nu_e$. An
alternative set of data is constituted by the decay rates and the measured
values of the $g_1/f_1$ ratios. Since the latter, however, contain less
experimental information, using the several angular coefficients instead of
$g_1/f_1$ provides a more sensitive test.

In order to extract $V_{us}$ several analyses can be performed under
different assumptions. The analysis done in Ref.~\cite{csw} neglects
the quadratic SB corrections in the vector form factor $f_1$ and accounts
for the larger effects in the axial form factor $g_1$ by using the measured
values of the $g_1/f_1$ ratios of the above processes, except for $\Xi^-\to
\Sigma^0 e^- \overline \nu_e$. The matrix element $V_{us}$ is then
extracted separately in each decay and the results are combined to obtain
the value $V_{us}=0.2250\pm 0.0027$, which is in good agreement with the
unitarity requirement.

A similar analysis has been done in Ref.~\cite{rfm04}, this time using the
decay rates and the angular correlation coefficients and performing a
global fit to the data of the five decays. By first assuming the validity of
exact SU(3) symmetry, the analysis yields $V_{us}=0.2238\pm 0.0019$, with a
$\chi^2/{\rm dof}$ of around 2.5. This high value of $\chi^2$ may signal the
presence of not negligible SB corrections. One can proceed further and
study the effects upon $V_{us}$ of the leading form factors when SB
corrections are taken into account. If one fixes $f_1$ at their SU(3)
symmetry values and incorporates first-order SB in $g_1$ one obtains
$V_{us}=0.2230\pm 0.0019$. Next, incorporating SB effects in both $f_1$ and
$g_1$ yields $V_{us}=0.2199 \pm 0.0026$ \cite{rfm04}. This analysis
also finds that SB corrections to $f_1$ increase their magnitudes over
their SU(3) symmetric predictions by up to 7\% and that corrections to $g_1$
are consistent with expectations. This latter $V_{us}$, although
in good agreement with the one quoted by the Particle Data Group, does not
help in improving the unitarity test. Another interesting finding is that
the incorporation of more refined SB corrections systematically reduces the
value of $V_{us}$ from its SU(3) symmetric prediction, rather than
increasing it to better satisfy unitarity. Of course an increase of $V_{us}$
can be achieved with a symmetry breaking pattern such that $f_1/f_1^{\rm
SU(3)} <1$ \cite{don,sch,kra}, but current data do not seem to favour this
trend \cite{rfm04}.

We are now in a position to make some statements. The first one is that
the assumption of exact SU(3) symmetry to compare theory and experiment in
HSD is questionable due to the poor fits it produces. The second one is the
fact that deviations from the exact SU(3) limit, in particular for
the case of the leading axial form factor $g_1$, are indeed important to
reliably determine $V_{us}$ from HSD, which can rival in precision with the
one from $K_{l3}$ decays. Nevertheless more work, theoretical and
experimental, is required in the next future to be conclusive about $V_{us}$
from HSD. Particularly, recent techniques in lattice QCD \cite{latt} allow
to measure form factors with great accuracy, thus determining SB corrections
in a model-independent manner.

\subsection{Determination of $|V_{us}|$ and $m_s$ from hadronic $\tau$ decays}

Already over a decade ago, it was realized {\cite{BNP92}} that
hadronic $\tau$ decays provide a very clean testing ground for low
energy QCD. For example, analysing the non strange $\tau$ spectral function
\cite{Barate:1998uf} leads to a determination of the QCD coupling
$\alpha_s$ that is competitive with the world average. Also, in view
of the increasingly precise data on the strange spectral function from
ALEPH \cite{ALEPH99}, OPAL \cite{OPAL04} and CLEO \cite{CLEO03},
studying flavor breaking observables allows for a determination of the
strange quark mass, another fundamental QCD parameter. The first
analyses of this kind \cite{PP99,CKP98,KM00,CDGHPP01} were plagued
with large uncertainties in the scalar contributions due to a bad
behavior of the associated perturbation series. This problem was
circumvented \cite{GJPPS03} by substituting the theoretical
expressions with phenomenological counterparts, thereby reducing the
theoretical uncertainties significantly. Along with this procedure, it
was realized that the main uncertainty in the strange quark mass
determination was now arising from the uncertainty in the CKM
parameter $|V_{us}|$, which suggests {\cite{GJPPS03,Gamiz:2004ar}} to
turn the analysis around and determine $|V_{us}|$.  

The basic objects for a QCD analysis of hadronic $\tau$ decays are
the two point correlation functions for vector 
$V^{\mu}_{ij} \equiv \overline{q}_i \gamma^\mu q_j$ and axial-vector
$A^{\mu}_{ij} \equiv \overline{q}_i \gamma^\mu \gamma_5 q_j$ currents.
Both correlation functions contain (pseudo)scalar and
(axial) vector contributions.
Then, the ratio of hadronic to leptonic $\tau$ decay rates
\begin{equation}
R_\tau \equiv 
\frac{\Gamma\left[\tau^- \to {\rm hadrons} (\gamma)\right]}
{\Gamma\left[ \tau^- \to e^- \overline{\nu}_e \nu_\tau (\gamma)\right]},
\end{equation}
as well as higher moments of the invariant mass distribution
\begin{equation}
R_\tau^{(k,l)} \equiv {\dis \int^{M_\tau^2}_0}
{\rm d} s \left(1-\frac{s}{M_\tau^2}\right)^k \, 
\left( \frac{s}{M_\tau^2} \right)^l \, \frac{{\rm d} R_\tau}{{\rm d} s}
\end{equation}
can be calculated in an operator product expansion:
\begin{eqnarray}
R_\tau^{(k,l)} \equiv N_c S_{\rm EW} 
\Big\{ (|V_{ud}|^2 + |V_{us}|^2) \,  \left[ 1 + \delta^{(k,l)(0)}\right]
 \nonumber \\
 + {\dis \sum_{D\geq2}} \left[ |V_{ud}|^2 \delta^{(k,l)(D)}_{ud}
+ |V_{us}|^2 \delta^{(k,l)(D)}_{us} \right] \Big\} \, . 
\end{eqnarray}
Here, $S_{EW}=1.0201\pm0.0003$ summarizes electroweak radiative corrections,
while explicit expressions for the $\delta^{(k,l)(D)}_{ij}$ and
additional theoretical information can be found in
\cite{PP99} as well as references therein.
The dominant contribution to  $R_{\tau}$ is purely perturbative and summarized
in $\delta^{(k,l)(0)}$, while higher dimensional contributions that
also depend on the flavor content are suppressed. The most important
of these suppressed terms are proportional to $m_s^2$ and $m_s \langle
\overline q q \rangle $. To reduce perturbative uncertainties one
conveniently discusses the flavor breaking observable 
\begin{equation}
\delta R^{(k,l)}_\tau \equiv
\frac{R^{(k,l)}_{\tau,V+A}}{|V_{ud}|^2}-
\frac{R^{(k,l)}_{\tau,S}}{|V_{us}|^2}  = 
N_c \, S_{EW} \, {\dis \sum_{D\geq 2}}
 \left[ \delta^{(k,l)(D)}_{ud}
-\delta^{(k,l)(D)}_{us}\right] \, , \nonumber
\end{equation}
where the definition of the flavour-dependent moments 
$R^{(k,l)}_{\tau,V+A}$ and $R^{(k,l)}_{\tau,S}$
can be found in Ref.~\cite{GJPPS03}. 
In this expression, the main parametric uncertainties arise only from $m_s$
and $|V_{us}|$, so that one could ideally
determine both parameters simultaneously from the experimental
analysis of several moments (see~\cite{GJPPS03} for more details). In
absence of such an analysis, we proceed as in {\cite{Gamiz:2004ar}}
and begin by determining $|V_{us}|$ from the moment with the smallest dependence on
$m_s$, i.e. the (0,0) moment. Using $m_s( 2 {\rm GeV})=(95\pm20)$ MeV
(in the the $\overline{MS}$ scheme), 
in agreement with recent sum rule and lattice calculations, 
one finds $\delta R_{\tau,\rm th}^{(0,0)}=0.218\pm0.026$, and, with
\begin{equation}
\label{Vusdet}
|V_{us}|^2 = 
\frac{R^{(0,0)}_{\tau,S}}
{\frac{\dis R^{(0,0)}_{\tau,V+A}}
{\dis |V_{ud}|^2}-\delta R^{(0,0)}_{\tau,{\rm th}}}
\end{equation}
one obtains
\begin{equation}
\label{value}
|V_{us}|=0.2208\pm0.0033_{\rm exp}\pm 0.0009_{\rm th}
= 0.2208 \pm 0.0034,
\end{equation}
where we have used $R^{(0,0)}_{\tau,V+A}=3.469\pm0.014$  and
$R^{(0,0)}_{\tau,S}=0.167\pm0.0050$ from \cite{OPAL04}. The most important feature
of this determination is the small theoretical uncertainty, the reason
for which can nicely be seen from Eq.~(\ref{Vusdet}): the large cancellations
between strange and non strange channels lead to a small value of
$\delta R^{(0,0)}_{\tau,{\rm th}}$, so that the main sensitivity is to
the experimental input. It is also interesting to check whether this
value of $|V_{us}|$ is compatible with unitarity:
using the value $|V_{ud}| = 0.9739\pm0.0003$ (see section~2),
one finds that Eq.~(\ref{eq:unitarity})
is violated only at the 1.8 $\sigma$ level.

In the next step, one uses the value of $|V_{us}|$ thus obtained and
determines the strange quark mass from higher moments. One finds the
values given in Table~\ref{table1}, where the individual sources of
the uncertainties are given in {\cite{Gamiz:2004ar}}.

\begin{table}[t]
\centering
\begin{tabular}{cccc}
\hline 
Moment & (2,0) & (3,0) & (4,0) \\
\hline 
$m_s(M_\tau)$ MeV & $93.2^{+34}_{-44}$ & $86.3^{+25}_{-30}$ 
& $79.2^{+21}_{-23}$ \\
\hline
\end{tabular}
\caption{Results for $m_s(M_\tau)$ extracted from the different
moments. \label{table1}}
\end{table}

The weighted average of the strange mass values
obtained for the different moments give
\begin{equation}
\label{ms}
m_s(M_\tau^2)= 84 \pm 23 \, {\rm MeV} \,  
\Rightarrow  m_s(2{\rm GeV})= 81 \pm 22 \, {\rm MeV} \ ,
\end{equation}
in good agreement with the average given above, while a more detailed
comparison can be found in \cite{Gamiz:2004ar}.
The values of $m_s$ display a monotonous $k$ dependence, which has
been significantly reduced with the new OPAL data, mainly due to the new
value for the $B(\tau^-\to K^- \pi^+ \pi^- \nu)$ branching ratio. 

In summary, the $\left|V_{us}\right|$ value obtained from $\tau$
decays is beginning to become competitive with the standard
determination from $K$ decays and is given in Eq.~(\ref{value}).
In particular, due to the fact that the
largest amount of the uncertainty is still experimental, one can expect
that the better data samples from BaBar and Belle will reduce
this uncertainty significantly. In view of these perspectives it will
be important to have the possibility to determine both $m_s$ and
$|V_{us}|$ simultaneously. An analysis of this type is
underway. Additionally, further, precise data will clarify whether the moment
dependence of the $m_s$ results is indeed a purely experimental issue,
and thereby allow a consistency check of the whole analysis.

\section{Future prospects on $V_{cs}$ and $V_{cd}$}

\subsection{Theoretical developments}
As far as the extraction of $V_{cs}$ and $V_{cd}$ is concerned,
the most interesting recent developments are the 
lattice QCD calculations of the matrix elements 
for semileptonic and leptonic $D_{(s)}$ decays.
With the advent of dynamical simulations with light sea quarks, decays to $K$
and $\pi$ mesons are favoured because the final-state particles are
stable to strong decay. The headline results have come from
simulations using $2{+}1$ flavours of improved staggered light quarks
where the common light ($u$, $d$) quark mass is in the range $m_s/8 <
m_\mathrm{light} < 3m_s/4$. The ``fourth root trick'' used in generating
these dynamical configurations has not been completely justified
theoretically, but on the other hand it has not so far failed any
test. There is a requirement to deal with the unphysical ``tastes''
introduced, but this has been addressed by the development of
staggered chiral perturbation theory
(S$\chi$PT)~\cite{Aubin:2004xd,ab-sl-stag}, allowing the subtraction
of discretisation effects arising from light quark taste violations.

\subsubsection{Lattice Results for Semileptonic $D$ Decays}

For semileptonic $D$ decays to a light final state pseudoscalar or
vector meson, $P$ or $V$, the squared momentum transfer is $q^2 =
m_D^2+m_{P,V}^2 - 2 m_D E$ where $E$ is the $P$ or $V$ energy in the
$D$ rest frame. In lattice simulations, the entire range, $0 \leq q^2
\leq (m_D-m_{P,V})^2 = \qsqmax$, can be accessed while keeping the
initial and final state meson spatial momenta small enough to avoid
large discretisation effects.

Two form factors, $f_+(q^2)$ and $f_0(q^2)$ are needed to describe the
vector current matrix element for semileptonic decays to a light
pseudoscalar, but only $f_+(q^2)$ is needed for the decay rate if the
lepton mass can be ignored. Lattice QCD gives results for both form
factors and this can be helpful since they can be fit simultaneously
imposing the constraint $f_+(0)=f_0(0)$.

The Fermilab-MILC-HPQCD collaboration has reported results for
semileptonic $D\to K$ and $D\to\pi$ decays using $2{+}1$ flavours of
improved staggered quarks~\cite{Aubin:2004ej,Okamoto:2004xg}. The
charm quark is implemented using the Fermilab
method~\cite{El-Khadra:1996mp}, while the calculations are so far
performed at one lattice spacing. Indeed, the dominant systematic
effect comes from the heavy quark discretisation error, estimated from
the mismatch of the continuum and lattice heavy quark effective
theories~\cite{Kronfeld:2000ck,Kronfeld:2003sd} and comprising $7\%$
of the $10\%$ total systematic error in the form factors.

\begin{figure}
\parbox{0.4\textwidth}{%
\includegraphics[width=\hsize]{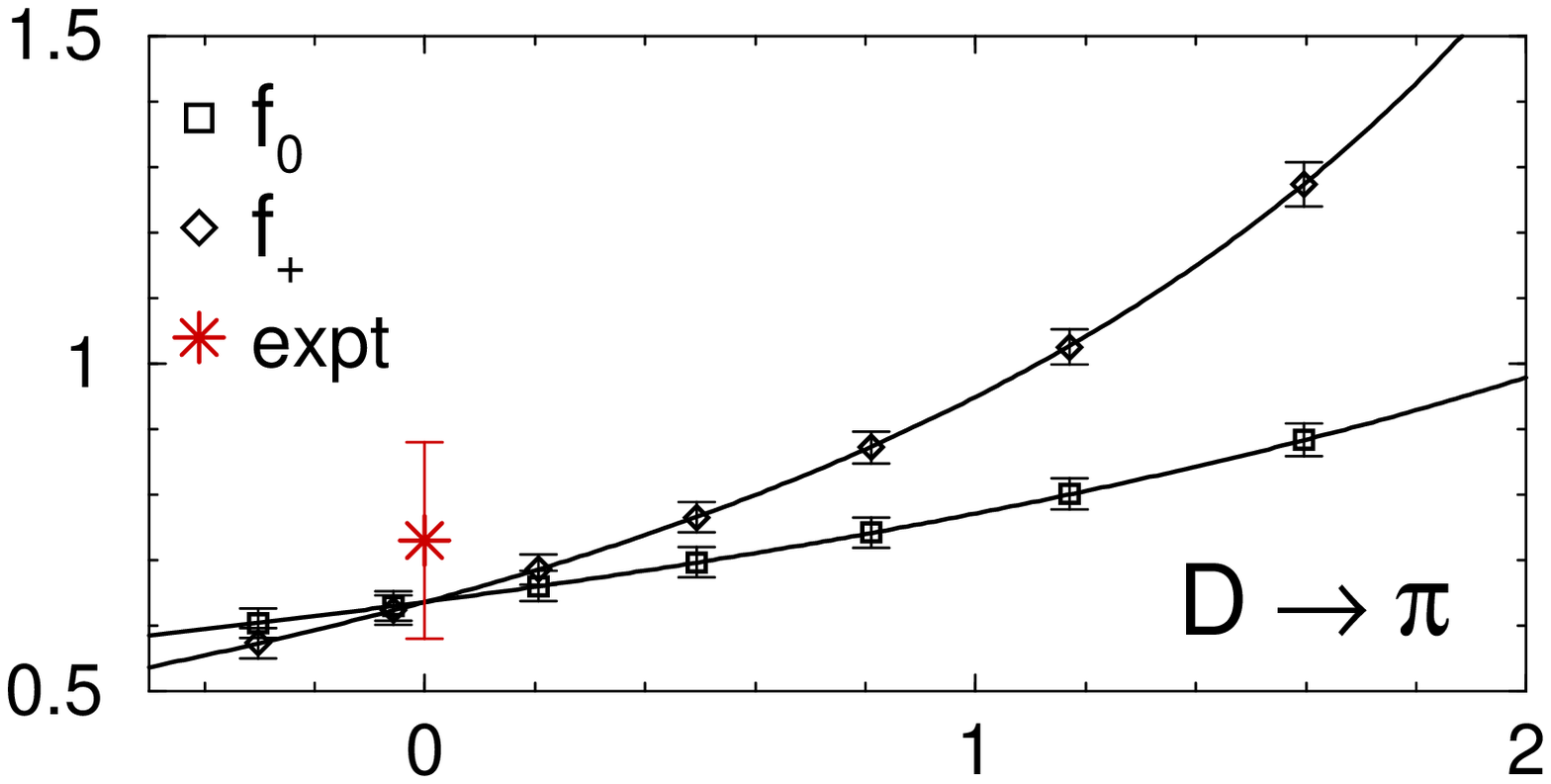}
\includegraphics[width=\hsize]{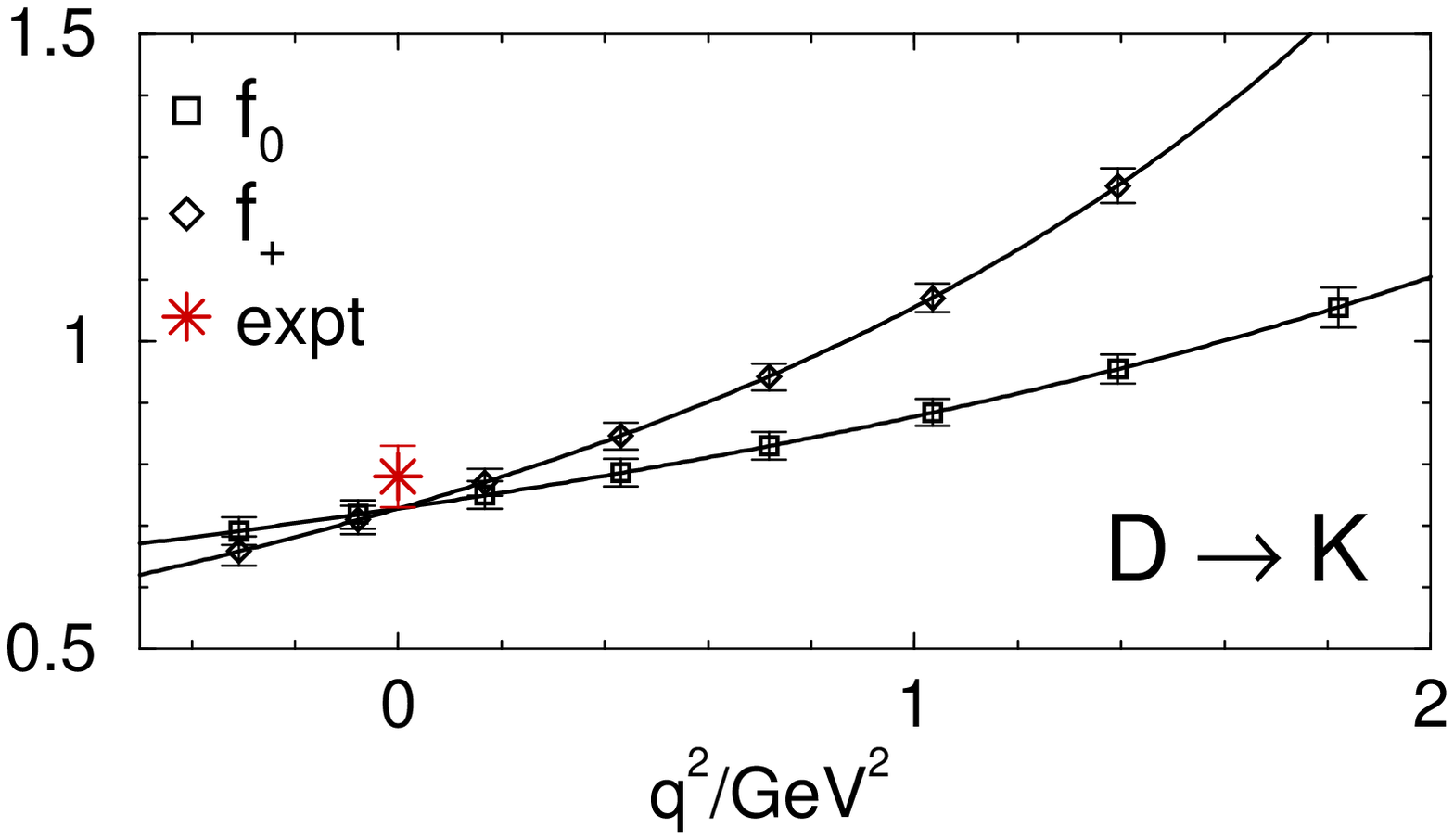}}
\hfill
\parbox{0.475\textwidth}{%
\newdimen\unit
\unit=0.095\textwidth
\psset{unit=5\unit,xunit=5\unit,yunit=3\unit}
\input{lattice-focus-compare.ps}}
\caption{Left: Fermilab-MILC-HPQCD results~\cite{Aubin:2004ej} for
  $D\to\pi$ and $D\to K$ semileptonic decay form factors. The
  experimental points at $q^2=0$ are from BES~\cite{Ablikim:2004ej}.
  Right: comparison of FNAL-MILC-HPQCD calculation~\cite{Aubin:2004ej}
  of $f_+$ for $D\to Kl\nu$ with FOCUS data~\cite{Link:2004dh}. The
  green curve is a fit to the BK parametrisation using the lattice
  data. The green triangles show $q^2$ values where the fixed-$E$
  chiral extrapolation ends up. The green curve and points use the
  central values of the fitted BK parameters. The blue points are from
  FOCUS.}
\label{fig:fnal-dtopi-dtok}
\end{figure}

It is convenient to perform the chiral extrapolations at fixed pion
energy, $E$. To do this, the form factors for different light quark
masses are fit to the Becirevic-Kaidalov (BK)
parametrisation~\cite{BK-param}, which satisfies $f_+(0)=f_0(0)$,
obeys heavy quark scaling relations near zero recoil ($\qsqmax$) and
at $q^2=0$ and contains a $D^*_{(s)}$ pole in $f_+$. They are then
interpolated to a common set of fixed values of $E$ and extrapolated
using the S$\chi$PT expressions~\cite{Aubin:2004xd,ab-sl-stag}. After
this, $f_{+,0}$ are recovered and a final BK fit is made to extend the
results to the full kinematic range\footnote{Extrapolations have also
  been made using the S$\chi$PT expressions without the intermediate
  BK fits. The results are consistent but
  noisier~\cite{Kronfeld-Tsukuba2004}.}. The output form factors are
shown in figure~\ref{fig:fnal-dtopi-dtok} and compared to FOCUS
results~\cite{Link:2004dh} for $f_+^{D\to K}$. Results for $f_+(0)$,
compared to quenched calculations, are:
\[
\def\arraystretch{1.3}
\begin{array}{rll}
               & N_f=0   & N_f=2{+}1  \\
f_+^{D\to\pi}(0) & 0.62(7) & 0.64(3)(6) \\
f_+^{D\to K}(0)  & 0.71(8) & 0.73(3)(7) \\
f_+^{D\to\pi}(0)/f_+^{D\to K}(0) &
                         & 0.87(3)(9)
\end{array}
\]
The quenched results are an average
from~\cite{Bowler:1994zr,Allton:1994ui,Bhattacharya:1994fp,Bhattacharya:1995zc,wup97,Abada:2000ty},
with an added $10\%$ error incorporated to allow for unquenching and
lack of a continuum extrapolation.

For decays to light vector mesons, only quenched results are
available. The most recent results for $D\to K^* l\nu$ come from the
SPQcdR collaboration~\cite{spqcdr-b2rho-2002}. There are four form
factors, $V$ and $A_{0,1,2}$, but $A_0$ is not needed for the decay
rate. $A_2$ is the least well determined, but since it is common to
quote values at $q^2=0$, $A_2(0)$ can be determined from $A_0(0)$ and
$A_1(0)$. Combining the SPQcdR results at their finer lattice spacing
with previous results
from~\cite{Bowler:1994zr,Allton:1994ui,Bhattacharya:1995zc,wup97},
gives:
\[
V(0)   = 1.17(14), \qquad
A_1(0) = 0.69(7), \qquad
A_2(0) = 0.58(8).
\]
SPQcdR fit their form factors to pole/dipole ans\"atze (consistent
with heavy quark scaling laws) and then integrate to
find~\cite{spqcdr-b2rho-2002} (at their finer lattice spacing):
\[
\Gamma(D^-\to K^{*0}l\nu)/|V_{cs}^2| = 0.062(15)\,\mathrm{ps}^{-1}
\]

\subsubsection{Lattice Results for $\fds$ and $\fd$}

\begin{figure}
\begin{center}
\small
\definecolor{midgrey}{gray}{0.5}
\definecolor{lightgrey}{gray}{0.9}
\definecolor{purple}{rgb}{0.8,0,0.8}
\definecolor{dkgreen}{rgb}{0,0.8,0}
\definecolor{dkcyan}{rgb}{0,0.7,0.8}
\definecolor{dkred}{rgb}{0.8,0,0}
\def\result#1#2(#3,#4,#5){%
\rput[l](75,#1){#2}
\psline(#4,#1)(#5,#1)
\rput(#4,#1){\psline(0,-0.15)(0,0.15)}
\rput(#5,#1){\psline(0,-0.15)(0,0.15)}
\pscircle[fillstyle=solid](#3,#1){0.12}
}
\newdimen\unit
\unit=0.035\hsize
\psset{arrowinset=0,linewidth=0.03}
\psset{unit=\unit,xunit=0.06\unit,yunit=0.7\unit}
\begin{pspicture}(70,-1.8)(340,22)
\psframe[linestyle=none,fillstyle=solid,fillcolor=lightgrey](234,0)(300,21)
\rput[t](200,-1.2){$200$}
\rput[t](250,-1.2){$250$}
\rput[t](300,-1.2){$300$}
\rput[tl](340,-1.2){$f_{D_s}/\mev$}
\multirput(190,0)(10,0){16}{\psline(0,-1)(0,-0.7)\psline(0,22)(0,21.7)}
\multirput(200,0)(50,0){3}{\psline(0,-1)(0,-0.5)\psline(0,22)(0,21.5)}
\psset{linecolor=dkred,fillcolor=dkred}
\result1{FNAL-MILC-HPQCD 2005~\cite{Aubin:2005ar}}(249,232.7,265.3)
\result2{Wingate et al 2004~\cite{Wingate:2003gm}}(290,336,244)
\rput[l](70,3.3){$N_f=2{+}1$}
\psset{linecolor=dkgreen,fillcolor=dkgreen}
\result5{CPPACS 2005~\cite{Kayaba:2004hu} error stat only}(267,257,278)
\result6{MILC 2002~\cite{milc-fnal-stat-nf2-2002}}(241,215,270)
\result7{CPPACS 2001~\cite{cppacs-fnal-quen-nf2-2001}}(267,217,317)
\rput[l](70,8.3){$N_f=2$}
\psset{linecolor=dkcyan,fillcolor=dkcyan}
\result{10}{J\"uttner--Rolf 2003~\cite{Juttner:2003ns}}(252,243,261)
\result{11}{de Divitiis et al 2003~\cite{deDivitiis:2003wy}}(240,233,247)
\psset{linecolor=purple,fillcolor=purple}
\result{12.5}{Lellouch--Lin 2001~\cite{ukqcd-ll-rel-quen-2001}}(236,219.9,254.8)
\result{13.5}{UKQCD 2001~\cite{ukqcd-rel-quen-2001}}(229,216.6,252)
\result{14.5}{Becirevic et al 1999~\cite{ape-rel-quen-1999}}(231,219,245.4)
\psset{linecolor=blue,fillcolor=blue}
\result{16}{MILC 2002~\cite{milc-fnal-stat-nf2-2002}}(223,205.3,241.7)
\result{17}{CPPACS 2001~\cite{cppacs-fnal-quen-nf2-2001}}(250,204.6,295.4)
\result{18}{MILC 1998~\cite{milc-rel-stat-quen-1998}}(210,197.2,241.5)
\result{19}{El Khadra et al 1998~\cite{fnal-fnal-quen-1998}}(213,197.4,230.8)
\result{20}{JLQCD 1998~\cite{jlqcd-fnal-quen-1998}}(223,204.6,243.4)
\rput[l](70,21.3){$N_f=0$}
\end{pspicture}
\end{center}
\caption{History of $\fds$ calculations in lattice QCD (adapted from
  plot shown by Hashimoto~\cite{Hashimoto:2004hn} at ICHEP04). The
  shaded background shows the Particle Data Group~\cite{PDG2004}
  experimental world average $\fds = (267\pm33)\mev$, with $|V_{cs}|$
  as input.}
\label{fig:fds-history}
\end{figure}
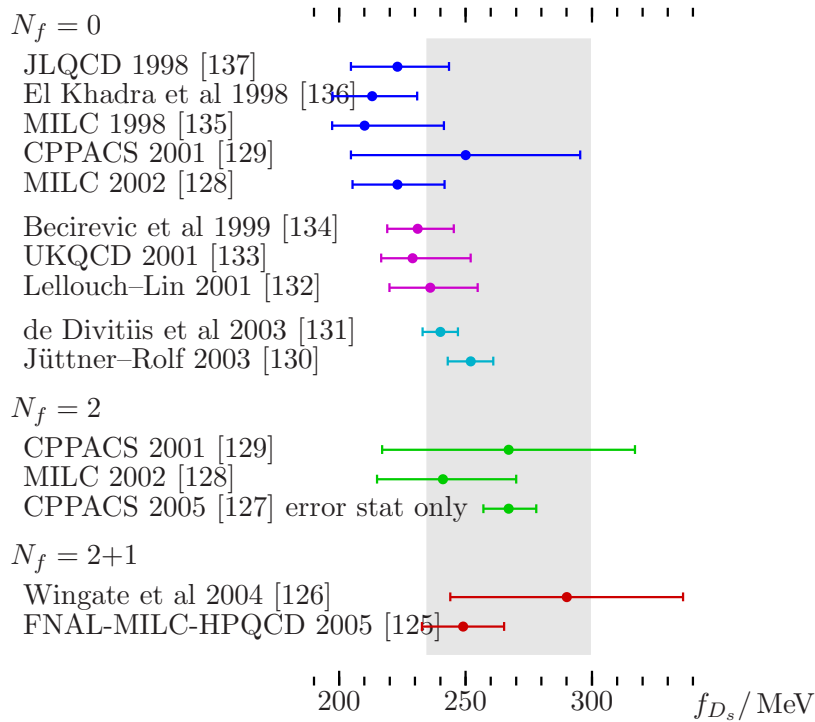

Recent quenched lattice results for $D$-meson decay constants have
concentrated on controlling all systematic errors save for quenching
itself~\cite{Juttner:2003ns,deDivitiis:2003wy}. Attention has now
shifted to simulations with dynamical
quarks~\cite{Wingate:2003gm,Kayaba:2004hu,Bernard:2004kz,Simone:2004fr}.
Figure~\ref{fig:fds-history} shows results from lattice calculations
of $\fds$ published from 1998 onwards.

Dynamical simulations with two clover flavours of light quark
(although with $m_\mathrm{sea} > m_s/2$) are in
progress~\cite{Kayaba:2004hu}. Simulations using $2{+}1$ flavours of
improved staggered quarks currently allow the lightest sea quark
masses to be reached. In one case~\cite{Wingate:2003gm}, the heavy
quark is treated using NRQCD and the extrapolation to the $D_s$ meson
tends to magnify errors. The Fermilab-MILC-HPQCD collaboration treat
the charm quark directly and use a large set of valence and sea quark
masses to help control the chiral
extrapolation~\cite{Simone:2004fr,Aubin:2005ar}. A fit using
S$\chi$PT~\cite{Aubin:2004xd} allows discretisation errors from
staggering to be removed. For $\fds$, the valence quark mass is
interpolated to the strange mass, and then the sea quark mass is
extrapolated to the up--down mass. In contrast, for $\fd$, the valence
and sea masses are set equal and extrapolated together. The results
are~\cite{Aubin:2005ar}:
\[
\def\arraystretch{1.2}
\begin{array}{rcl}
\fds &=& 249(3)(16)\mev \\
\fd  &=& 201(3)(17)\mev
\end{array}
\]


\subsubsection{Sumrule Results}

Sumrules are generally less stable for $D$ mesons than for $B$ mesons:
there are larger higher order operator and perturbative corrections.
For the semileptonic decays of $D$ mesons to $\pi$ or $K$ mesons,
Khodjamirian is updating the lightcone sumrule
predictions~\cite{kho-lcsr2005} using an updated value for the charm
quark mass, $m_c = 1.46\pm0.1\gev$~\cite{PDG2004}, and new
Gegenbauer moments for the kaon and pion distribution
amplitudes~\cite{Khodjamirian:2004ga}. The preliminary results,
compared to the summary~\cite{Khodjamirian:2003rf} from the 2003 CKM
workshop, are:
\[
\def\arraystretch{1.3}
\begin{array}{rll}
 & \mbox{CKM03~\cite{Khodjamirian:2003rf}} & \mbox{2005~\cite{kho-lcsr2005}}\\
f^{D\to\pi}_+(0) & 0.65(11) & 0.61(11) \\
f^{D\to K}_+(0) & 0.78(11)  & 0.79(14)(8) \\
f^{D\to\pi}_+(0)/f^{D\to K}_+(0) & & 0.77(4)(8)
\end{array}
\]
The second error in $f_+^{D\to K}$ arises from variation in the
strange quark mass, $m_s = 130\mp20\mev$.

For the decay constants, the summary from the 2003 CKM workshop still
stands~\cite{Khodjamirian:2003rf}:
\[
\fd = 200\pm20\mev,
\qquad
\fds/\fd = 1.11 \mbox{\ to\ } 1.27
\]

\subsection{Recent experimental results on charm decays}   

Precision measurements of semileptonic charm decay rates 
and form factors are a principal goal of the CLEO-c program 
at the Cornell Electron Storage Ring~(CESR)\cite{cleoc}.
We review herein measurements with the first CLEO-c data of the absolute 
branching fractions of $D^0$ decays to $K^- e^+ \nu_e$, $\pi^- e^+ \nu_e$
and  $K^{*-} e^+ \nu_e$,
as well as of $D^+$ decays to  $\bar{K}^0 e^+ \nu_e$, $\pi^0 e^+ \nu_e$, 
$\bar{K}^{*0} e^+ \nu_e$ and $\rho^0 e^+ \nu_e$, including
the first observations and absolute branching fraction 
measurements of $D^0 \rightarrow \rho^- e^+ \nu_e$ and $D^+ \rightarrow 
\omega e^+ \nu_e$~\cite{cleoc-neutral_semilep,cleoc-charged_semilep},
and compare them to recent measurements from other experiments and 
theoretical predictions. We also review the prospects for measuring 
semileptonic form factors and the CKM matrix elements $V_{cs}$ and $V_{cd}$ 
with the full CLEO-c data set.

The data for this analysis were collected by the CLEO-c 
detector at the $\psi (3770)$ resonance, about 40~MeV above 
the $D \bar{D}$ pair production threshold.
A description of the CLEO-c detector is provided in 
Ref.~\cite{cleoc-neutral_semilep} and references therein. 
The data sample consists of an integrated luminosity of 
55.8~pb$^{-1}$ and includes about 160,000 $D^+ D^-$
and 200,000 $D^0 \bar{D}^0$ events.

The technique for these measurements was first applied by the Mark III
collaboration~\cite{MkIII} at SPEAR.
Candidate events are selected by reconstructing a $\bar{D}^0$~($D^-$), 
called a tag, in a hadronic final state.
The absolute branching fractions of $D^0$~($D^+$) semileptonic decays are 
then  measured by their reconstruction in the system recoiling from the tag.
Tagging a $\bar{D}^0$~($D^-$) meson in a $\psi(3770)$ decay provides a 
$D^0$~($D^+$) with known four-momentum, allowing a semileptonic decay 
to be reconstructed with no kinematic ambiguity, even though the neutrino 
is undetected.

Tagged events are selected based on two variables: $\Delta E
\equiv E_{D} - E_{\rm beam}$,  the difference between the energy 
of the $D^{-}$ tag candidate ($E_{D}$) and the beam energy ($E_{\rm beam}$), 
and the beam-constrained mass $M_{\rm bc} \equiv \sqrt{ E_{\rm beam}^2/c^4 - 
|\vec{p}_{D}|^2/c^2}$, where $\vec{p}_{D}$ is the measured momentum of 
the $D^{-}$ candidate~\cite{cleoc-Dtagging}. 
If multiple candidates are present in the same tag mode, one candidate per 
$\bar{D}^0$ or $D^-$ is chosen using $\Delta E$, and 
the yields in each tag modes are obtained from fits to 
the $M_{\rm bc}$ distributions. The data sample comprises approximately 
60,000~(32,000) neutral~(charged) tags, reconstructed
in eight neutral~(six charged) $D$ meson hadronic final 
states~\cite{cleoc-neutral_semilep,cleoc-charged_semilep}, respectively.

After a tag is identified, the positron
and a set of hadrons recoiling against the tag is 
searched for.
(Muons are not used as $D$ semileptonic decays at the $\psi(3770)$
produce low momentum leptons for which the CLEO-c muon
identification is not efficient.)
The efficiency for positron identification rises from about
$50\%$ at 200~MeV/$c$ to 95\% just above 300~MeV/$c$ and is roughly 
constant thereafter, while the positron fake rate from charged pions 
and kaons is approximately 0.1\%~\cite{cleoc-neutral_semilep}.
Candidates for $\pi^0$ are identifid from photon pairs, each 
having an energy of at least 30 MeV, with invariant mass 
within 3.0$\sigma$ ($\sigma \sim 6$~${\rm MeV}/c^2$)  
of the known $\pi^0$ mass. The $K^0_S$ candidates are formed from pairs of 
oppositely-charged and vertex-constrained tracks having an invariant mass 
within 12 MeV/$c^2$ $( \sim 4.5 \sigma )$ of the known $K^0_S$ mass. 
$\bar{K}^{*-}$/$\rho^-$/$\bar{K}^{*0}$/$\rho^0$ candidates are formed 
from  ($K^-$ and $\pi^0$) or ($K^0_S$ and $\pi^-$) / ($\pi^-$ and $\pi^0$) / 
($K^-$ and $\pi^+$) / ($\pi^-$ and $\pi^+$)
combinations and require an invariant mass within 100~MeV/$c^2$ and 
150~MeV/$c^2$ for the  $K^*$ and $\rho$ from the expected mean values.  
The reconstruction of $\omega \rightarrow \pi^+ \pi^- \pi^0$ candidates 
is achieved by combining three pions, requiring an invariant mass within
20~MeV/$c^2$ of the known mass, and demanding that the charged pions 
do not satisfy interpretation as a ${K}^0_S$.

The tag and the semileptonic decay are then combined,
if the event includes no tracks other than those of the tag
and the semileptonic candidate. Semileptonic decays are identified using 
the variable $U \equiv E_{\rm miss} - |\vec{p}_{\rm miss}|c$, 
where $E_{\rm miss}$ and $\vec{p}_{\rm miss}$ are the missing energy and 
momentum of the $D$ meson decaying semileptonically. If the decay products 
of  the semileptonic decay have been correctly identified, 
$U$ is expected to be zero,  since only a neutrino is undetected. 
The $U$ distribution has $\sigma \sim 10~{\rm MeV}$. 
(The width varies by mode and is larger for modes 
with $\pi^0$ mesons.) To remove multiple candidates in each 
semileptonic mode, one combination is chosen per tag mode, based 
on the proximity of the invariant masses of the $K_S^0$,  $\bar{K}^{*}$, 
$\rho$, $\pi^0$, or $\omega$ candidates to their expected masses.

The new results for the branching ratios of various  semileptonic  modes 
are reported in  Table~\ref{table2}.
For each mode the yield 
is determined from a fit to
its $U$ distribution.  The backgrounds are generally small and arise mostly 
from misreconstructed semileptonic decays with correctly reconstructed tags. 
The absolute branching fractions are then  determined 
using $ {\cal B}={N_{\rm signal} / \epsilon N_{\rm tag}}$,
where $N_{\rm signal}$ is the number of fully reconstructed $D \bar{D}$
events obtained by fitting the $U$ distribution, $N_{\rm tag}$ is the number 
of events with a reconstructed tag, and $\epsilon$ is the effective efficiency 
for detecting the semileptonic decay in an event with an identified tag.

\begin{table}[t]
\begin{center}
\begin{tabular}{ l c c c}
\hline \hline
 Mode &  Yield   &  $\mathcal{B}$ (\%)   &   $\mathcal{B}$ (\%) (PDG) \\
\hline
$D^0 \rightarrow K^- e^+ \nu_e$      & $1311 \pm 37$ & $3.44 \pm 0.10 \pm 0.10$ &  $3.58 \pm 0.18$ \\
$D^0 \rightarrow \pi^- e^+ \nu_e$    & $116.8 \pm 11.2$ & $0.26 \pm 0.03 \pm 0.01$ &  $0.36 \pm 0.06$ \\
$D^0 \rightarrow K^{*-} e^+ \nu_e$   & $219.3 \pm 15.6$  & $2.16 \pm 0.15 \pm 0.08$ &  $2.15 \pm 0.35$ \\
$D^0 \rightarrow \rho^{-} e^+ \nu_e$ & $31.1 \pm 6.3$ & $0.19 \pm 0.04 \pm 0.01$ &  --- \\
$D^+ \rightarrow  \bar{K}^0 e^+ \nu_e $ & $545 \pm 24 $
                                      & $ 8.71 \pm 0.38 \pm 0.37 $  &  $6.7  \pm 0.9$        \\
$D^+ \rightarrow  \pi^0 e^+ \nu_e  $  & $63.0 \pm 8.5$
                                      & $ 0.44 \pm  0.06 \pm 0.03 $  &  $0.31 \pm 0.15$   \\
$D^+ \rightarrow  \bar{K}^{*0} e^+ \nu_e  $ & $422 \pm 21$   
                                      & $5.56 \pm 0.27 \pm 0.23 $  &  $5.5  \pm 0.7$ \\
$D^+ \rightarrow  \rho^0 e^+ \nu_e  $ & $27.4 \pm 5.7$   
                                      & $ 0.21 \pm 0.04 \pm 0.01 $ & $0.25 \pm 0.10$ \\
$D^+ \rightarrow  \omega e^+ \nu_e  $ & $ 7.6^{+3.3}_{-2.7}$
                                      &  $ 0.16^{+0.07}_{-0.06} \pm 0.01 $ &  ---  \\
\hline \hline
\end{tabular}
\end{center}
\caption{Signal yields and branching fractions as recently obtained at CLEOc
         vs. the PDG results~\cite{PDG2004}.
         The first uncertainty is statistical and the second systematic in 
         the third column, and statistical or total in the other columns.
         The yield~($\mathcal{B}$) for $D^0 \rightarrow  K^{*-} e^+ \nu_e$ 
         is summed~(averaged) over the two $K^{*-}$ submodes of $K^- \pi^0$
         and $K^0_S \pi^-$. Signal yields for $D^0 \rightarrow \rho^- e^+ \nu_e$ 
         and $D^+ \rightarrow \omega e^+ \nu_e$ are significant and represent 
         the first observations of these modes. \label{table2} }
\end{table}

The largest contributors to the systematic uncertainty are associated 
with the tracking efficiency, the $\pi^0$ and $K_S^0$ reconstruction efficiencies, 
the extraction of $D$ tag yields, the positron and hadron identification 
efficiencies, background shapes and normalizations, imperfect knowledge of 
form factors, and the simulation of final state radiation. 
The non-resonant contribution in the $D \rightarrow K^* e \nu_e$ reconstruction
and the associated systematic uncertainty are accounted for as described 
in Ref.~\cite{cleoc-neutral_semilep,cleoc-charged_semilep}.
The total systematic uncertainty ranges from 2.8\% to
7.8\% according to the mode. Most systematic uncertainties are measured 
in data and will be reduced with a larger data set.

The widths of the isospin conjugate exclusive semileptonic decay modes 
of the $D^0$ and $D^+$ are related by isospin invariance of the hadronic
current.  The ratio $\frac{\Gamma(D^0 \rightarrow 
K^- e^+ \nu_e)} {\Gamma(D^+ \rightarrow \bar{K}^0 e^+ \nu_e)}$  is expected
to be unity. The world average value is $1.35 \pm 0.19$~\cite{PDG2004}.
Using the new results and the lifetimes of the $D^0$ and
$D^+$~\cite{PDG2004}, one obtains:  $\frac{\Gamma(D^0 \rightarrow K^-e^+\nu_e)} 
{\Gamma(D^+ \rightarrow \bar{K}^0 e^+ \nu_e)} = 1.00 \pm 0.05 {\rm (stat)} 
\pm 0.04 {\rm (syst)}.$
The result is consistent with unity and with two recent
less precise results: a measurement from BES II using the same
technique~\cite{BESII_ratio} and an indirect measurement from
FOCUS~\cite{FOCUS_ratio,FOCUS_ksmunu}. The ratios of isospin
conjugate decay widths for other semileptonic decay modes are
given in Ref.~\cite{cleoc-charged_semilep} and are all consistent
with isospin invariance.


As the data are consistent with isospin invariance, the precision
of each branching fraction can be improved by averaging the $D^0$ 
and $D^+$ results for isospin conjugate pairs. The isospin-averaged 
semileptonic decay widths, with correlations among systematic uncertainties
taken into account, are given in Table~\ref{decayWidths}.

\begin{table}[t]
\begin{center}
\begin{tabular}{ l  c }
\hline \hline
 Decay Mode \hspace{10mm} & \hspace{5mm} $\Gamma$~($10^{-2}\times{\rm ps}^{-1}$) \hspace{5mm} \\
\hline
$D \rightarrow  K \; e^+ \nu_e  $     & $8.38 \pm   0.20 \pm  0.23 $  \\
$D^0 \rightarrow  \pi^- e^+ \nu_e $   & $0.68 \pm   0.05 \pm  0.02 $  \\
$D \rightarrow  K^* \; e^+ \nu_e $    & $5.32 \pm   0.21 \pm  0.20 $  \\
$D^0 \rightarrow  \rho^- e^+ \nu_e$   & $0.43 \pm   0.06 \pm  0.02 $  \\
\hline \hline
\end{tabular}
\end{center}
\caption{Isospin-averaged semileptonic 
         decay widths with
         statistical and systematic uncertainties.
         For Cabibbo-suppressed modes,
         the isospin average is calculated for the $D^0$
         using $\Gamma(D^0) = 2 \times \Gamma(D^+)$. \label{decayWidths} }
\end{table}

The ratio of decay widths for $D \rightarrow \pi  e^+ \nu$ and $D
\rightarrow K e^+ \nu$ provides a test of the LQCD charm semileptonic 
rate ratio prediction~\cite{unquenched_LQCD}. 
Using the results in Table~\ref{decayWidths}, one finds $\frac{\Gamma(D^0 
\rightarrow \pi^- e^+ \nu)} {\Gamma(D \rightarrow K e^+ \nu)} = (8.1 
\pm 0.7 {\rm (stat)}  \pm 0.2 {\rm (syst)} ) \times 10^{-2}$, consistent
with LQCD and two recent results~\cite{pikenu_cleo,pikenu_focus}. 
Furthermore, the ratio $\frac{\Gamma(D
\rightarrow K^* e^+ \nu)} {\Gamma(D \rightarrow K e^+ \nu)}$ is
predicted to be in the range 0.5 to 1.1 (for a compilation see
Ref.~\cite{FOCUS_ratio}). Using the isospin averages in Table~\ref{decayWidths}, 
one finds $\frac{\Gamma(D \rightarrow K^* e^+ \nu)} {\Gamma(D
\rightarrow K e^+ \nu)} = 0.63 \pm 0.03 {\rm (stat)} \pm 0.02 {\rm
(syst)}$.


The modest in size data sample used in this analysis produced 
the most precise measurements to date of the absolute branching 
fractions for all modes in Table~\ref{table2}. The analysis
already provides stringent tests of the theory. This precision 
is consistent with the expected performance of CLEO-c. 
CESR is running to collect a much larger $\psi(3770)$ data sample.  

The future CLEO-c data sample~(3.0~fb$^{-1}$) will allow  measurements 
of the $q^2$ dependence~($f_+(q^2)$) and, assuming the unitarity of the 
CKM matrix, the absolute normalization~($f_+(0)$) of form factors at 
a percent level~($\mathcal{O}(1\%)$) in both $D \rightarrow K e^+ \nu_e$
and $D \rightarrow \pi e^+ \nu_e$~\cite{cleoc}. 
For $D \rightarrow K^* e^+ \nu_e$ and $D \rightarrow \rho e^+ \nu_e$, 
form factors $A_1(q^2)$, $A_2(q^2)$, and $V(q^2)$ can be measured 
to a few percent~\cite{cleoc}. These measurements of form factors 
(and other precision measurements in the $D$, $B$, $\Upsilon$ and
$\psi$ systems) constitute calibration and validation data 
for LQCD and other theoretical approaches.

Theoretical predictions for form factors are required for measurements of 
the CKM matrix elements. In charm decays, the modes
$D^0 \rightarrow K^- e^+ \nu_e$ and  $D^0 \rightarrow \pi^- e^+ \nu_e$ are 
important as they are the simplest for both theory and experiment. 
Assuming future theoretical uncertainties on the form factors in
these decays of 3.0\% and using the anticipated uncertainties for the absolute 
decay rates of $D^0 \rightarrow K^- e^+ \nu_e$ and  $D^0 \rightarrow \pi^- e^+ \nu_e$ 
from CLEO-c of 1.2\% and 1.5\%, respectively, the following uncertainties on $|V_{cs}|$ and 
$|V_{cd}|$ are within reach: $|\sigma(V_{cs})/V_{cs}| \approx 1.6\%$ and
$|\sigma(V_{cd})/V_{cd}| \approx 1.7\%$~\cite{cleoc}.

\section{Conclusions}
As discussed in section~\ref{sect:Vud},  at present 
the determination of $|V_{ud}|$ is largely dominated by 
super-allowed, $0^{+}\rightarrow0^{+}$ nuclear beta decays.
Taking into account the new theoretical analysis of radiative 
corrections in Ref.~\cite{marciano13}, and the new global 
fit by Savard {\em et al.}~\cite{savard}, leads to 
\begin{equation}
|V_{ud}|= 0.97377 \pm 0.00027~.
\label{eq:Vud_fin}
\end{equation}
Hopefully, in the near future a competitive independent 
information on $|V_{ud}|$ will be extracted from 
the neutron beta decay, once the experimental 
situation on the neutron lifetime will be clarified.

Concerning $V_{us}$, at present the most reliable determination 
is the one obtained by means of $K_{\ell 3}$ decays --see Eq.~(\ref{eq:fin_kl3})-- 
employing the Leutwyler-Roos value of $f_+(0)$  \cite{Leutwyler:1984je},
which is supported by several recent lattice 
results \cite{becirevic,Okamoto2,Okamoto1,JLQCD,RBC}:
\be
 |V_{us}|_{K_{\ell 3}} = 0.2261 \pm 0.0021  
  \qquad  \left[ f_+(0) = 0.961 \pm 0.008 \right]~.
\ee
Of comparable accuracy is the value extracted from 
the $\Gamma(K_{\mu 2})/\Gamma(\pi_{\mu2})$ ratio,
employing the updated lattice result of $F_K/F_\pi$ \cite{MILC2}:
\be
 |V_{us}|_{K_{\mu 2}} =  0.2245^{+0.0011}_{-0.0031} 
 \qquad  \left[ F_K/F_\pi =1.198 \pm 0.003^{+0.016}_{-0.005} \right]~. 
\ee
The average of these two values leads to 
\be
 |V_{us}|_{K~{\rm decays}} =  0.225 \pm 0.001~,
\label{eq:Vus_fin}
\ee
which can be considered as the final 
global estimate of $V_{us}$.

Using the values in Eqs.~(\ref{eq:Vud_fin}) and (\ref{eq:Vus_fin})
the unitarity relation in Eq.~(\ref{eq:unitarity}) 
is satisfied within $1 \sigma$ at the $10^{-3}$ level:
\begin{equation}
\Delta = 1 - \left(|V_{ud}|^2+|V_{us}|^2+|V_{ub}|^2\right) = (1 \pm 1 ) \times 10^{-3}~.
\end{equation}

\section*{Acknowledgements}  
This work was supported by the U.S.~DOE grant
No.~DE-AC02-98CH1086, by the U.S.~NSF grant No.~NSF PHY-0245068, 
and by the E.U. IHP-RTN program, contract 
No.~HPRN-CT-2002-00311 (EURIDICE).


\begin{thebibliography}{99}
\bibitem{FOCUS_ksmunu} FOCUS Collaboration measures
$\Gamma(D^+ \rightarrow \bar{K}^{*0} \mu^+\nu_{\mu})/$
$\Gamma(D^+ \rightarrow \bar{K}^0 \mu^+ \nu_{\mu})$
and $\Gamma(D^+ \rightarrow \bar{K}^{*0} \mu^+\nu_{\mu})/$
$\Gamma(D^+ \rightarrow K^- \pi^+ \pi^+)$. Using
the world average value of
${\cal B} (D^+ \rightarrow K^- \pi^+ \pi^+)$ they extract
$ {\cal B} (D^+ \rightarrow \bar{K}^0 \mu^+ \nu_{\mu})$
and find a value much larger than that in the PDG.
\end{thebibliography}
\end{document}